\documentclass[12pt]{article}
\usepackage{amsmath,amssymb}
\usepackage{graphicx}
\graphicspath{ {./images/} }
\usepackage[font=small,labelfont=bf]{caption}
\usepackage{amsmath}
\usepackage{braket}
\usepackage{hyperref}
\usepackage{pgfplots}
\pgfplotsset{compat=1.15}
\usepackage{mathrsfs}
\usepackage{pgf,tikz}
\usetikzlibrary{arrows}

\usepackage{float}
\usepackage{color, soul}
\sethlcolor{yellow}
\newcommand {\be}{\begin{equation}}
 \newcommand {\ee}{\end{equation}}
 \newcommand {\bea}{\begin{array}}
 
 \newcommand {\eea}{\end{array}}

\textwidth=6.5in \hoffset=-.75in \voffset=-1in \numberwithin{equation}{section}
\numberwithin{figure}{section}

\def\0{{(0)}}
\def\1{{(1)}}
\def\2{{(2)}}

\def\<{\langle }
\def\>{\rangle }
\def\[{\left[}
\def\]{\right]}

\begin{document}
\begin{titlepage}

\vskip1cm
\begin{center}
{~\\[140pt]{ \LARGE {\textsc{The reflected entropy in the GMMG/GCFT flat holography    }}}\\[-20pt]}
\vskip2cm

\end{center}
\begin{center}
{Mohammad Reza Setare * \footnote{E-mail: rezakord@ipm.ir}\hspace{1mm} ,
Meisam Koohgard * \footnote{E-mail: m.koohgard@modares.ac.ir}\hspace{1.5mm}\hspace{2mm} \\
{\small {\em  {* Department of Science,\\
 Campus of Bijar, University of Kurdistan, Bijar, Iran }
  }}}\\
\end{center}
\begin{abstract}
We extend the reflected entropy to the bipartite state in a two dimensional Galilean conformal field theory ($GCFT_2$) which is dual to the asymptotically flat spacetime described by the generalized minimal massive gravity (GMMG). To this end we consider the Renyi entanglement entropy formula that can be related to the reflected entropy by some manipulations of the replica technique. In a limited case, we find the reflected entropy is twice the minimal entanglement wedge cross section (EWCS) and this is an evidence on the relation between the reflected entropy and the EWCS in the flat holography scenario.                 \end{abstract}
\vspace{1cm}


\end{titlepage}

\section{Introduction}	
The computation of the entanglement measure is a necessary ingredient in the quantum information theories. A viable measure in the gauge/gravity holography \cite{hol01,hol02,hol03,hol04,hol05} is the entanglement entropy that on a bipartite pure state $\rho_{AB}=\ket\psi\bra\psi$, can be computed by the von Neumann entropy. The state $\rho_{AB}$ could be defined on a Hilbert space $\mathcal{H}_A\otimes\mathcal{H}_B$. The von Neumann entropy of the reduced density matrix $\rho_A=\mathrm{Tr}_B\rho_{AB}$ has the following relation \cite{EE03,EE04,EE05,EE08,EE09,EE10}
\begin{equation}\label{EE01}
  S(A)=-\mathrm{Tr}\rho_A\ln\rho_A.
\end{equation}

Using the Ryu-Takayanagi formula as follows \cite{EE08,EE09,EE10}
\begin{equation}\label{EE-RT}
  S(A)=\frac{\mathcal{A(\mathcal{M})}}{4G_N},
\end{equation}
it is possible to correspond a codimension-2 surface in the bulk of $AdS_3$ geometry to the entanglement entropy of a $CFT_2$ at the boundary. $A(\mathcal{M})$ is the area of the minimal surface in the bulk known as the RT surface that is homologous to the entangling surface at the boundary. However, the entanglement entropy is not a viable measure of the classical and quantum correlations for the bipartite mixed states or the states with more than two intervals at the boundary \cite{EE10}. It is needed to find other correlation measures for the multipartite or mixed bipartite systems.

A new holographic object to describe the bipartite mixed states has been developed by the minimal entanglement wedge cross section (EWCS) that can be computed as the extremal surface that can split the entanglement wedge into two areas \cite{EW01,EW02,EW03}. The entanglement wedge as a distinct region from the causal wedge \cite{EW01,EW02,EW03,CW01,CW02,CW03} is a bounded region in the bulk by the RT surfaces and the bipartite intervals A and B and is dual to the reduced density matrix of the CFT. From another point of view, the minimal entanglement wedge cross section has been introduced as a holographic dual of the
entanglement of purification (EoP) \cite{EP01,EP02,EP03,EP04,EP05,EP06} that is defined as a minimal area by some special manipulations in the bulk. In more explicit words, the entanglement of purification $E_P$ for a bipartite state can be defined on a doubled Hilbert space  $\mathcal{H_{A}}\otimes\mathcal{H_{B}}\otimes\mathcal{H_{A^*}}\otimes\mathcal{H_{B^*}}$ as follows
\begin{equation}\label{EP01}
  E_P(A:B)=_{\ket\psi _{AA^*BB^*}}min~S(AA^*)
\end{equation}
where the minimization can be done over all purifications $\ket\psi_{AA^*BB^*}$. It is difficult to prove the equality between $E_p$ and EWCS, because of some considerations on the boundary computations. It should be noted that the EWCS in the bulk dual of the bipartite mixed states in the CFT has been related to other measures of the entanglement:  odd entanglement entropy \cite{OE} , reflected entropy \cite{RE01,RE02,RE03,RE04} and logarithmic negativity \cite{Ku01}.

Due to the difficulty mentioned above, the reflected entropy has been extended in \cite{RE01} to be another boundary quantity dual to EWCS. The reflected entropy has the following relation with the EWCS
\begin{equation}\label{RE01}
  S_R(A:B)=\frac{\mathcal{A}[\mathcal{M}_R]}{4G_N}=2E_W(A:B),
\end{equation}
where $\mathcal{A}[\mathcal{M}_R]$  is the area of an algorithmically constructed region in the bulk as the reflected minimal surface. The advantage of $S_R$ over the $E_P$ is that it has no minimizations over the purifications and it is easier to be calculated.

The construction for the reflected entropy in \cite{RE01} has been developed for the multipartite scenario at the boundary in \cite{RE03,RE04}. The authors in \cite{RE04} have extended the construction of the minimal reflected surface to the multipartite states and they have utilized the replica technique in the $AdS_3/CFT_2$ scenario.\\

In this paper, we use the replica technique developed in \cite{RE04} in the flat holography case where we consider the two dimensional Galilean conformal field theory ($GCFT_2$) at the boundary of the holography. The two dimensional Galilean conformal field theories ($GCFT_2$) were proposed as the non-relativistic version of the corresponding conformal field theories in two dimensional spacetime by using the Inonu-Wigner contraction of the $CFT_2$ algebra \cite{GC01,GC02,GC03,GC04}. The non-relativistic flat holography between the bulk gravity and boundary
quantum field theory has been extended in \cite{holG01}.\\

The three dimensional Generalized Minimal Massive Gravity (GMMG) \cite{GMMG01} could be an option for the bulk description of the two dimensional Galilean conformal algebra (2d GCA) with some asymmetric central charges. GMMG model is a theory that avoids the bulk-boundary clash and as a Minimal Massive Gravity (MMG) \cite{MMG01} has some positive energy excitations around the vacuum that are the maximally $AdS_3$. The GMMG's central charges are positive in the dual CFT. In contrast to GMMG, the Topologically Massive Gravity (TMG) \cite{TMG01,TMG02} and the cosmological extension of New Massive Gravity \cite{NMG} that are constructed previously with local degrees of freedom
in three dimensional spacetime could not avoid the aforementioned clash.

In the GMMG/GCFT duality, we find that the reflected entropy is twice the EWCS. This result is important from different aspects. First of all, we have found a proper structure for the reflected entropy in the flat holography. Secondly, since the GMMG model for the gravity is more completed than the TMG model, it is more important to find quantities related to the quantum entanglement. In addition, compared to the AdS/CFT scenario investigated in \cite{RE04}, we have worked on the models that have two central charges $c_L$ and $c_M$ and there is no chiral symmetry. This feature is also present in the central charges in the $GCA_2$ algebra found in the GCFT theory at the boundary.

The paper is organized as follows. In section \ref{sec:2}, we give the definition of the reflected entropy for the bipartite mixed states. Given the construction developed in \cite{RE04}, we present the special construction that is needed to the Renyi entropy that is related to the $S_R$ using the replica trick.   In section \ref{sec:3} and the subsections, we consider the $S_R$ for the two disjoint intervals in the $GCFT_2$ in three cases: the vacuum, a finite temperature and a finite sized system. For the holography of the reflected entropy of the corresponding cases, we consider the GMMG describing the bulk geometries. To find the reflected entropy, we extend a special replica technique to construct the related Renyi entropy. In the following, we make a connection between the 4-point correlation functions developed in the Renyi entropy and the conformal blocks are related to the monodromy analysis \cite{Ku01,mon01,Seng01,Seng02}. The result of the 4-point correlation functions is an important ingredient to find the result of the Renyi entropy. The n-point correlation functions in the large central charge limit are dominated by the leading Galilean conformal blocks. We make a connection with our previous work on a conjecture on the minimal entanglement wedge cross section in the GMMG/GCFT scenario in \cite{GMMG02} to find the conformal blocks relations in three cases mentioned above. In section \ref{sec:4} and the subsections, we develop the approach extended in section \ref{sec:3}, for the two adjacent intervals at three different cases in the GMMG/GCFT flat holography. We conclude in section \ref{sec:5} with a summary of our results.

\section{The reflected entropy}\label{sec:2}
We review some general futures of the reflected entropy using the papers \cite{RE01,RE02,RE03,RE04,RE05} in this section. The reflected entropy has been introduced in \cite{RE01}. Given a mixed state $\rho_{AB}$, the canonical purification of the state can be defined in a doubled Hilbert space as follows \cite{RE01}
\begin{equation}\label{Pu01}
  \ket{\sqrt{\rho_{AB}}}\in(\mathcal{H}_A\otimes\mathcal{H}_{A^*})\otimes(\mathcal{H}_B\otimes\mathcal{H}_{B^*})
  \equiv\mathcal{H}_{AA^*BB^*}.
\end{equation}
It can be found that the purification could double the Hilbert space as follows \cite{RE03}
\begin{equation}\label{Pu02}
  \rho_{AB}=\sum_{i} p_i\ket{\psi_i}\bra{\psi_i}~~~ \to~~~\ket{\sqrt{\rho_{AB}}}=\sum_{i}{\sqrt{p_i}}\ket{\psi_i}_{AB}
  \ket{\psi_i}_{A^*B^*}^*,
\end{equation}
where $\ket\psi_i$ is an orthonormal basis for the bipartite state $\rho_{AB}$. Another relation that is a sign for the purification can be found as follows \cite{RE01}
\begin{equation}\label{Pu03}
  \mathrm{Tr}_{A^*B^*}\ket{\sqrt{\rho_{AB}}}\bra{\sqrt{\rho_{AB}}}=\rho_{AB}.
\end{equation}

The reflected entropy $S_R$ is the von Neumann entropy across $AA^*$ as follows \cite{RE01}
\begin{equation}\label{RE201}
  S_R(A:B)=S(AA^*)_{\ket{\sqrt{\rho_{AB}}}},
\end{equation}
where the von Neumann entropy has the following relation
\begin{equation}\label{vN01}
  S(AA^*)=-\mathrm{Tr}_{AA^*} \rho_{AA^*}\log \rho_{AA^*}.
\end{equation}

Since the complementary part of $AA^*$ is $BB^*$ part of the doubled Hilbert space, $\rho_{AA^*}$ has the following relation
\begin{equation}\label{Pu04}
  \rho_{AA^*}=\mathrm{Tr}_{BB^*}\ket{\sqrt{\rho_{AA^*}}}\bra{\sqrt{\rho_{AA^*}}}
\end{equation}
where we have used the relation (\ref{Pu03}).

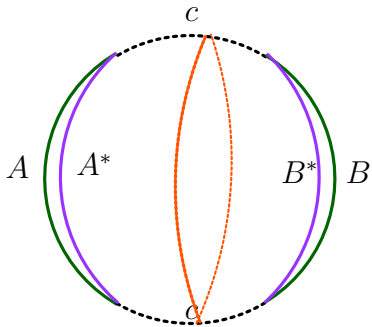
\begin{figure}[!h]
\centering
\captionsetup{width=.8\linewidth}
\definecolor{ffvvqq}{rgb}{1,0.3333333333333333,0}
\definecolor{zzttff}{rgb}{0.6,0.2,1}
\definecolor{qqwuqq}{rgb}{0,0.39215686274509803,0}
\begin{tikzpicture}[line cap=round,line join=round,>=triangle 45,x=1cm,y=1cm]
\clip(-0.6,0.5) rectangle (4.62,5.6);
\draw [shift={(2,3)},line width=1.2pt,dotted]  plot[domain=1.017764882644256:2.1129857367217753,variable=\t]({1*1.9037857022259623*cos(\t r)+0*1.9037857022259623*sin(\t r)},{0*1.9037857022259623*cos(\t r)+1*1.9037857022259623*sin(\t r)});
\draw [shift={(2,3)},line width=1.2pt,color=qqwuqq]  plot[domain=2.1129857367217753:4.179086797655315,variable=\t]({1*1.937937047481161*cos(\t r)+0*1.937937047481161*sin(\t r)},{0*1.937937047481161*cos(\t r)+1*1.937937047481161*sin(\t r)});
\draw [shift={(2,3)},line width=1.2pt,dotted]  plot[domain=4.179086797655315:5.259951216324687,variable=\t]({1*1.9276929216034384*cos(\t r)+0*1.9276929216034384*sin(\t r)},{0*1.9276929216034384*cos(\t r)+1*1.9276929216034384*sin(\t r)});
\draw [shift={(2,3)},line width=1.2pt,color=qqwuqq]  plot[domain=-1.0232340908548991:1.017764882644256,variable=\t]({1*1.920833152566875*cos(\t r)+0*1.920833152566875*sin(\t r)},{0*1.920833152566875*cos(\t r)+1*1.920833152566875*sin(\t r)});
\draw [shift={(2.48,3.02)},line width=1.2pt,color=zzttff]  plot[domain=2.304957322788926:3.996940202028023,variable=\t]({1*2.209072203437452*cos(\t r)+0*2.209072203437452*sin(\t r)},{0*2.209072203437452*cos(\t r)+1*2.209072203437452*sin(\t r)});
\draw [shift={(1.42,3.02)},line width=1.2pt,color=zzttff]  plot[domain=-0.8100845054530481:0.7916873886501558,variable=\t]({1*2.2917242417009946*cos(\t r)+0*2.2917242417009946*sin(\t r)},{0*2.2917242417009946*cos(\t r)+1*2.2917242417009946*sin(\t r)});
\draw (1.78,5.42) node[anchor=north west] {$c$};
\draw (1.78,1.46) node[anchor=north west] {$c$};
\draw (0.34,3.48) node[anchor=north west] {$A^*$};
\draw (-0.6,3.44) node[anchor=north west] {$A$};
\draw (3.06,3.38) node[anchor=north west] {$B^*$};
\draw (3.92,3.38) node[anchor=north west] {$B$};
\draw [shift={(-2.7,3.24)},line width=0.8pt,dash pattern=on 1pt off 1pt,color=ffvvqq]  plot[domain=-0.4253720916261683:0.32529123762539236,variable=\t]({1*5.248843552976105*cos(\t r)+0*5.248843552976105*sin(\t r)},{0*5.248843552976105*cos(\t r)+1*5.248843552976105*sin(\t r)});
\draw [shift={(6.94,2.9)},line width=1.2pt,dash pattern=on 1pt off 1pt,color=ffvvqq]  plot[domain=2.743446545765593:3.501066243946351,variable=\t]({1*5.140705026954927*cos(\t r)+0*5.140705026954927*sin(\t r)},{0*5.140705026954927*cos(\t r)+1*5.140705026954927*sin(\t r)});
\end{tikzpicture}

\caption{Canonical purification of the mixed bipartite state $\rho_{AB}$ by doubling the Hilbert space and gluing the $c$ parts of each circles. The red dashed line defines the reflected entropy.}
\label{fig201}
\end{figure}

The authors in \cite{RE02,RE04} have introduced a graph description of the reflected entropy . Given a pure state $\psi_{ABc}$ at the boundary, doubling the Hilbert space corresponds to double the circle on which the state is defined. The purified state $\ket{\sqrt{\rho_{AB}}}=\ket{\sqrt{Tr_c\ket{\psi}\bra{\psi}}}$ can be described by gluing the $c$ parts of the circles as depicted in (Fig.\ref{fig201})

The red curve in Fig.\ref{fig201} is twice the minimal entanglement wedge cross section (EWCS) that can separate two boundaries $AA^*$ and $BB^*$. The graph description of $S_R$ in Fig.\ref{fig201} can be found in the following relation \cite{RE04}
\begin{equation}\label{RE202}
  S_R(A:B)= Entanglement~entropy~of~the~Red~curve,
\end{equation}
where the red curve is an entangling surface in the space.

\begin{figure}[!h]
\centering
\captionsetup{width=.8\linewidth}

\definecolor{ffvvqq}{rgb}{1,0.3333333333333333,0}
\definecolor{qqqqff}{rgb}{0,0,1}
\definecolor{zzttff}{rgb}{0.6,0.2,1}
\definecolor{qqwuqq}{rgb}{0,0.39215686274509803,0}
\begin{tikzpicture}[line cap=round,line join=round,>=triangle 45,x=1cm,y=1cm]
\clip(-0.6,0.4) rectangle (4.6,5);
\draw [shift={(2,3)},line width=1.2pt,color=qqwuqq]  plot[domain=2.1129857367217753:4.179086797655315,variable=\t]({1*1.937937047481161*cos(\t r)+0*1.937937047481161*sin(\t r)},{0*1.937937047481161*cos(\t r)+1*1.937937047481161*sin(\t r)});
\draw [shift={(2,3)},line width=1.2pt,color=qqwuqq]  plot[domain=-1.0232340908548991:1.017764882644256,variable=\t]({1*1.920833152566875*cos(\t r)+0*1.920833152566875*sin(\t r)},{0*1.920833152566875*cos(\t r)+1*1.920833152566875*sin(\t r)});
\draw [shift={(2.48,3.02)},line width=1.2pt,color=zzttff]  plot[domain=2.304957322788926:3.996940202028023,variable=\t]({1*2.209072203437452*cos(\t r)+0*2.209072203437452*sin(\t r)},{0*2.209072203437452*cos(\t r)+1*2.209072203437452*sin(\t r)});
\draw [shift={(1.42,3.02)},line width=1.2pt,color=zzttff]  plot[domain=-0.8100845054530481:0.7916873886501558,variable=\t]({1*2.2917242417009946*cos(\t r)+0*2.2917242417009946*sin(\t r)},{0*2.2917242417009946*cos(\t r)+1*2.2917242417009946*sin(\t r)});
\draw (0.34,3.48) node[anchor=north west] {$A^*$};
\draw (-0.6,3.44) node[anchor=north west] {$A$};
\draw (3.06,3.38) node[anchor=north west] {$B^*$};
\draw (3.92,3.38) node[anchor=north west] {$B$};
\draw [shift={(2,6.14)},line width=2pt,color=qqqqff]  plot[domain=4.118175276584256:5.302821199032292,variable=\t]({1*1.7861690849412883*cos(\t r)+0*1.7861690849412883*sin(\t r)},{0*1.7861690849412883*cos(\t r)+1*1.7861690849412883*sin(\t r)});
\draw [shift={(2.04,0.02)},line width=2pt,color=qqqqff]  plot[domain=0.9491364045118077:2.223156653766968,variable=\t]({1*1.6483931569865244*cos(\t r)+0*1.6483931569865244*sin(\t r)},{0*1.6483931569865244*cos(\t r)+1*1.6483931569865244*sin(\t r)});
\draw [shift={(-2.18,3.18)},line width=0.8pt,dotted,color=ffvvqq]  plot[domain=-0.3439617799683763:0.2682235702530771,variable=\t]({1*4.482561237489698*cos(\t r)+0*4.482561237489698*sin(\t r)},{0*4.482561237489698*cos(\t r)+1*4.482561237489698*sin(\t r)});
\draw [shift={(7.58,3.02)},line width=0.8pt,dotted,color=ffvvqq]  plot[domain=2.9022844722195567:3.3808902421804308,variable=\t]({1*5.639386713870504*cos(\t r)+0*5.639386713870504*sin(\t r)},{0*5.639386713870504*cos(\t r)+1*5.639386713870504*sin(\t r)});
\draw (1.62,4.98) node[anchor=north west] {$\Gamma_{AB}$};
\draw (1.68,1.62) node[anchor=north west] {$\Gamma_{AB}$};
\draw (0.8,2.98) node[anchor=north west] {$M_{AB}$};
\draw (2.24,2.96) node[anchor=north west] {$M_{AB}$};
\end{tikzpicture}

\caption{The bulk dual of the canonical purification. The blue lines are the RT surfaces that tracing out $c$ corresponds to gluing these lines. The red dashed line corresponds to the holographic dual of the reflected entropy. }
\label{fig202}
\end{figure}
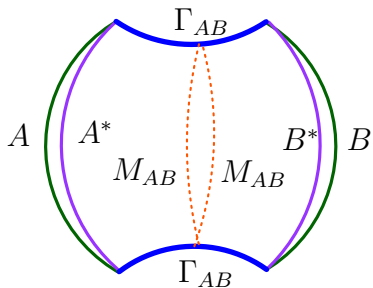

Given the gluing procedure to complete the purification at the boundary, it is needed to do the bulk dual of the procedure in a proper manner. Tracing out \emph{c} part in the bulk side of the duality corresponds to glue the Ryu-Takayanagi (RT) surfaces of the doubled bulk space (as depicted in Fig.\ref{fig202}). The RT surface $\Gamma_{AB}$ in the bulk is homologous to the mixed state $\rho_{AB}$ at the boundary. The area bounded by $A\cup B\cup\Gamma_{AB}$ is the entanglement wedge $M_{AB}$ and $\Sigma^{min}$ is a minimal surface that can split two entanglement wedges belonging to A
and B (the latter has been depicted by the red dashed lines in the Fig.\ref{fig202}).

Doubling the Hilbert space corresponds to double the entanglement wedge $M_{AB}$ and the canonical purification corresponds gluing the entanglement wedges along the RT surfaces. A replica technique by two indices \emph{n} and \emph{m} has been introduced to compute the reflected entropy $S_R(A:B)$ in \cite{RE01}. In this generalized replica technique , the $\rho_{AB}$ in (\ref{Pu02}) is converted to the following form
\begin{equation}\label{Rem01}
  \rho_{AB}^m=\sum_{i}p_i^m\ket{\psi_i}\bra{\psi_i}
\end{equation}
where using the Schmidt decomposition \cite{RE01,RE05,Hors} as follows
\begin{equation}\label{Rem02}
  \ket{\psi_i}=\sum_{a}\sqrt{l_i^a}\ket{a_i}_A\ket{a_i}_B,
\end{equation}
$ \rho_{AB}^m$ can be converted into the following form
\begin{equation}\label{Rem03}
   \rho_{AB}^m=\sum_{a,i,b}
    p_i^m\sqrt{l_i^al_i^b}\ket{a_i}_A\ket{a_i}_B\bra{b_i}_A\bra{b_i}_B.
\end{equation}

$\ket{a_i}_A$ and $\ket{a_i}_B$ are bases in $\mathcal{H}_A$ and $\mathcal{H}_B$, respectively. $l_i^a$ are nonnegative values with the following normalization relation
\begin{equation}\label{nor01}
  \Sigma_a l_i^a=1.
\end{equation}

Utilizing the Schmidt decomposition (\ref{Rem02}) in the purification (\ref{Pu02}) as follows
\begin{equation}\label{Rem04}
  \ket{\sqrt{\rho_{AB}}}=\sum_{a,i,b} \sqrt{p_il_i^a l_i^b},
  \ket{a_i}_A\ket{a_i}_B\ket{b_i}_{A^*}\ket{b_i}_{B^*}.
\end{equation}
$\rho_{AB}^{m/2}$ can be converted into the following form
\begin{equation}\label{Rem05}
  \ket{\rho_{AB}^{m/2}}=\sum_{a,i,b}p_i^{m/2}\sqrt{l_i^al_i^b}
  \ket{a_i}_A\ket{a_i}_B\ket{b_i}_{A^*}\ket{b_i}_{B^*}.
\end{equation}

By comparing (\ref{Pu02}) and (\ref{Rem05}), it is possible to define the purification of $\rho_{AB}^m$
as follows \cite{RE01,RE05}
\begin{equation}\label{Pum01}
 \ket{\psi_m}: =\frac{1}{\sqrt{\mathrm{Tr}\rho_{AB}^m}}\ket{\rho_{AB}^m},
\end{equation}
where its normalization has the following form
\begin{equation}\label{nor02}
  \mathrm{Tr}_{A^*B^*}\ket{\psi_m}\bra{\psi_m}=\frac{\rho_{AB}^m}{\mathrm{Tr}\rho_{AB}^m}.
\end{equation}

Using (\ref{RE201}), (\ref{RE202}) and (\ref{Pum01}), the von Neumann entropy is substituted by $S_n$ as follows \cite{RE01,RE05}
\begin{equation}\label{Rnm01}
  S_n(AA^*)_{\psi_m}:=\frac{1}{1-n}\log\mathrm{Tr}_{AA^*}\big(\rho_{AA^*}^{(m)}\big)^n,
\end{equation}
that is the $n^{th}$ Renyi entropy of the $\rho_{AA^*}^{(m)}$ with the following relation
\begin{equation}\label{Rnm02}
  \rho_{AA^*}^{(m)}: = \mathrm{Tr}_{BB^*}\ket{\psi_m}\bra{\psi_m}.
\end{equation}

Substituting (\ref{Rnm01}) in the following relation, the reflected entropy relation (\ref{RE01}) is obtained
\begin{equation}\label{REn01}
  \lim_{n,m\to 1}S_n(AA^*)_{\psi_m}=S_R(A:B).
\end{equation}

It is possible to compute the Renyi entropy using the following partition functions \cite{RE01}
\begin{equation}\label{Rnm03}
  Z_{n,m}:=\mathrm{Tr}_{AA^*}\big(\mathrm{Tr}_{BB^*}\ket{\rho_{AB}^{m/2}}\bra{\rho_{AB}^{m/2}}     \big)^n
\end{equation}

Using the partition functions (\ref{Rnm03}), the Renyi entropy (\ref{Rnm01}) could have the following form
\begin{equation}\label{Rnm04}
  S_n(AA^*)_{\psi_m}=\frac{1}{1-n}\log\frac{Z_{n,m}}{(Z_{1,m})^n}.
\end{equation}

\begin{figure}[!h]
\centering
\captionsetup{width=.8\linewidth}
\definecolor{ccwwff}{rgb}{0.8,0.4,1}
\definecolor{afeeee}{rgb}{0.6862745098039216,0.9333333333333333,0.9333333333333333}
\begin{tikzpicture}[line cap=round,line join=round,>=triangle 45,x=1cm,y=1cm]
\clip(4.8,-2.2) rectangle (9.2,3.2);
\draw [line width=1.2pt] (6,0)-- (5,-2);
\draw [line width=1.2pt] (6,0)-- (9,0);
\draw [line width=1.2pt] (9,0)-- (8,-2);
\draw [line width=1.2pt] (8,-2)-- (5,-2);
\draw [line width=1.2pt] (8.073333333333332,-0.6533333333333334)-- (7.6933333333333325,-1.4133333333333336);
\draw [line width=1.2pt] (7.293333333333333,-0.6333333333333334)-- (6.913333333333333,-1.3933333333333335);
\draw [line width=1.2pt] (8.073333333333332,-0.6533333333333345)-- (7.293333333333331,-0.6333333333333364);
\draw [line width=1.2pt] (6.913333333333333,-1.3933333333333335)-- (7.6933333333333325,-1.4133333333333336);
\draw [line width=1.2pt] (6.96,-0.64)-- (6.58,-1.4);
\draw [line width=1.2pt] (6.18,-0.62)-- (5.8,-1.38);
\draw [line width=1.2pt] (6.96,-0.64)-- (6.18,-0.62);
\draw [line width=1.2pt] (5.8,-1.38)-- (6.58,-1.4);
\draw [line width=1.2pt] (6,3)-- (5,1);
\draw [line width=1.2pt] (6,3)-- (9,3);
\draw [line width=1.2pt] (9,3)-- (8,1);
\draw [line width=1.2pt] (8,1)-- (5,1);
\draw [line width=1.2pt] (8.06,2.3466666666666662)-- (7.68,1.5866666666666667);
\draw [line width=1.2pt] (7.28,2.3666666666666667)-- (6.9,1.6066666666666667);
\draw [line width=1.2pt] (8.06,2.3466666666666596)-- (7.28,2.366666666666657);
\draw [line width=1.2pt] (6.9,1.6066666666666667)-- (7.68,1.5866666666666667);
\draw [line width=1.2pt] (6.96,2.36)-- (6.58,1.6);
\draw [line width=1.2pt] (6.18,2.38)-- (5.8,1.62);
\draw [line width=1.2pt] (6.96,2.36)-- (6.18,2.38);
\draw [line width=1.2pt] (5.8,1.62)-- (6.58,1.6);
\draw [line width=1.2pt,dotted,color=afeeee] (5.8,1.62)-- (5.90517857142857,1);
\draw [line width=1.2pt,color=ccwwff] (5.90517857142857,1)-- (6.18,-0.62);
\draw [line width=1.2pt,dotted,color=afeeee] (6.58,1.6)-- (6.681785714285715,1);
\draw [line width=1.2pt,color=ccwwff] (6.681785714285715,1)-- (6.96,-0.64);
\draw [line width=1.2pt,dotted,color=afeeee] (6.9,1.6066666666666667)-- (7.006527777777778,1);
\draw [line width=1.2pt,color=ccwwff] (7.293333333333331,-0.6333333333333364)-- (7.006527777777778,1);
\draw [line width=1.2pt,dotted,color=afeeee] (7.68,1.5866666666666671)-- (7.7830158730158745,1);
\draw [line width=1.2pt,color=ccwwff] (7.7830158730158745,1)-- (8.073333333333332,-0.6533333333333344);
\draw (6.033333333333336,2.2533333333333356) node[anchor=north west] {$A^*$};
\draw (7.1666666666666705,2.24) node[anchor=north west] {$B^*$};
\draw (6.033333333333336,-0.7466666666666624) node[anchor=north west] {$A$};
\draw (7.2066666666666706,-0.7333333333333291) node[anchor=north west] {$B$};
\end{tikzpicture}

\caption{The replica manifold of $\ket{\rho_{AB}^{m/2}}$ for $m=4$. Each sheet is in the different Hilbert space. The colored part corresponds to the tracing out in the definition of purification. }
\label{fig203}
\end{figure}
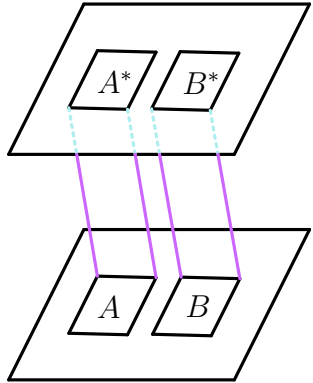

The authors in \cite{RE01} used a replica manifold for $Z_{n,m}$ in a 2d CFT with two disjoint intervals $A$ and $B$ where $n\in \mathrm{Z^+}$ and $m\in 2\mathrm{Z^+}$. For example in $m=4$, the replica manifold $\ket{\rho_{AB}^2}$ could be depicted by two sheets in Fig.\ref{fig203} in which we can see $\ket{\rho_{AB}^2}\in
\mathcal{H}_A\otimes\mathcal{H}_{A^*}\otimes\mathcal{H}_B\otimes\mathcal{H}_{B^*}$.

\begin{figure}[!h]
\centering
\captionsetup{width=.8\linewidth}
\definecolor{ffvvqq}{rgb}{1,0.3333333333333333,0}
\definecolor{ccwwff}{rgb}{0.8,0.4,1}
\definecolor{ccccff}{rgb}{0.8,0.8,1}
\definecolor{bcduew}{rgb}{0.7372549019607844,0.8313725490196079,0.9019607843137255}
\begin{tikzpicture}[line cap=round,line join=round,>=triangle 45,x=1cm,y=1cm]
\clip(-0.1,-9) rectangle (13,5);
\draw [line width=1.2pt] (1,0)-- (0,-2);
\draw [line width=1.2pt] (1,0)-- (4,0);
\draw [line width=1.2pt] (4,0)-- (3,-2);
\draw [line width=1.2pt] (3,-2)-- (0,-2);
\draw [line width=1.2pt] (3.0733333333333355,-0.6533333333333333)-- (2.6933333333333356,-1.413333333333333);
\draw [line width=1.2pt] (2.2933333333333352,-0.6333333333333337)-- (1.9133333333333353,-1.3933333333333335);
\draw [line width=1.2pt] (3.073333333333335,-0.6533333333333342)-- (2.293333333333334,-0.6333333333333367);
\draw [line width=1.2pt] (1.9133333333333353,-1.3933333333333335)-- (2.6933333333333356,-1.413333333333333);
\draw [line width=1.2pt] (1.96,-0.64)-- (1.58,-1.4);
\draw [line width=1.2pt] (1.18,-0.62)-- (0.8,-1.38);
\draw [line width=1.2pt] (1.96,-0.64)-- (1.18,-0.62);
\draw [line width=1.2pt] (0.8,-1.38)-- (1.58,-1.4);
\draw [line width=1.2pt] (1,3)-- (0,1);
\draw [line width=1.2pt] (1,3)-- (4,3);
\draw [line width=1.2pt] (4,3)-- (3,1);
\draw [line width=1.2pt] (3,1)-- (0,1);
\draw [line width=1.2pt] (3.06,2.3466666666666667)-- (2.68,1.5866666666666671);
\draw [line width=1.2pt] (2.28,2.366666666666667)-- (1.9,1.606666666666667);
\draw [line width=1.2pt] (3.06,2.3466666666666582)-- (2.28,2.3666666666666565);
\draw [line width=1.2pt] (1.9,1.606666666666668)-- (2.68,1.5866666666666736);
\draw [line width=1.2pt] (1.96,2.36)-- (1.58,1.6);
\draw [line width=1.2pt] (1.18,2.38)-- (0.8,1.62);
\draw [line width=1.2pt] (1.96,2.36)-- (1.18,2.38);
\draw [line width=1.2pt] (0.8,1.62)-- (1.58,1.6);
\draw (5.9336905083143465,2.1740997985618806) node[anchor=north west] {$A$};
\draw (7.021169709485077,2.14621571648058) node[anchor=north west] {$B$};
\draw (5.905806426233045,-0.893149230381199) node[anchor=north west] {$A^*$};
\draw (7.021169709485077,-0.893149230381199) node[anchor=north west] {$B^*$};
\draw [line width=1.2pt] (1,-6)-- (0,-8);
\draw [line width=1.2pt] (1,-6)-- (4,-6);
\draw [line width=1.2pt] (4,-6)-- (3,-8);
\draw [line width=1.2pt] (3,-8)-- (0,-8);
\draw [line width=1.2pt] (3.0733333333333355,-6.653333333333333)-- (2.693333333333335,-7.413333333333334);
\draw [line width=1.2pt] (2.2933333333333352,-6.633333333333334)-- (1.9133333333333353,-7.3933333333333335);
\draw [line width=1.2pt] (3.0733333333333355,-6.653333333333333)-- (2.2933333333333352,-6.633333333333334);
\draw [line width=1.2pt] (1.9133333333333442,-7.393333333333316)-- (2.693333333333344,-7.413333333333316);
\draw [line width=1.2pt] (1.96,-6.64)-- (1.58,-7.4);
\draw [line width=1.2pt] (1.18,-6.62)-- (0.8,-7.38);
\draw [line width=1.2pt] (1.96,-6.64)-- (1.18,-6.62);
\draw [line width=1.2pt] (0.8,-7.38)-- (1.58,-7.4);
\draw [line width=1.2pt] (1,-3)-- (0,-5);
\draw [line width=1.2pt] (1,-3)-- (4,-3);
\draw [line width=1.2pt] (4,-3)-- (3,-5);
\draw [line width=1.2pt] (3,-5)-- (0,-5);
\draw [line width=1.2pt] (3.06,-3.6533333333333338)-- (2.68,-4.413333333333333);
\draw [line width=1.2pt] (2.28,-3.6333333333333333)-- (1.9,-4.3933333333333335);
\draw [line width=1.2pt] (3.06,-3.6533333333333458)-- (2.28,-3.633333333333346);
\draw [line width=1.2pt] (1.9,-4.3933333333333335)-- (2.68,-4.413333333333333);
\draw [line width=1.2pt] (1.96,-3.64)-- (1.58,-4.4);
\draw [line width=1.2pt] (1.18,-3.62)-- (0.8,-4.38);
\draw [line width=1.2pt] (1.96,-3.64)-- (1.18,-3.62);
\draw [line width=1.2pt] (0.8,-4.38)-- (1.58,-4.4);
\draw (1.026092062005407,2.257752044805783) node[anchor=north west] {$A$};
\draw (2.113571263176138,2.1740997985618806) node[anchor=north west] {$B$};
\draw (0.8309034874363014,-0.7816129020559961) node[anchor=north west] {$A^*$};
\draw (2.0578030990135363,-0.8652651482998982) node[anchor=north west] {$B^*$};
\draw [line width=1.2pt] (6.010325333333335,-6.013874666666686)-- (5.010325333333335,-8.013874666666686);
\draw [line width=1.2pt] (6.010325333333335,-6.013874666666686)-- (9.010325333333338,-6.013874666666686);
\draw [line width=1.2pt] (9.010325333333338,-6.013874666666686)-- (8.010325333333334,-8.013874666666686);
\draw [line width=1.2pt] (8.010325333333334,-8.013874666666686)-- (5.010325333333335,-8.013874666666686);
\draw [line width=1.2pt] (8.083658666666668,-6.667208000000016)-- (7.703658666666668,-7.427208000000015);
\draw [line width=1.2pt] (7.303658666666669,-6.647208000000016)-- (6.923658666666669,-7.407208000000016);
\draw [line width=1.2pt] (8.083658666666658,-6.667208000000037)-- (7.303658666666669,-6.647208000000016);
\draw [line width=1.2pt] (6.923658666666704,-7.407207999999947)-- (7.703658666666668,-7.427208000000015);
\draw [line width=1.2pt] (6.970325333333335,-6.653874666666686)-- (6.590325333333335,-7.413874666666686);
\draw [line width=1.2pt] (6.190325333333336,-6.633874666666686)-- (5.810325333333336,-7.393874666666687);
\draw [line width=1.2pt] (6.970325333333324,-6.653874666666709)-- (6.190325333333336,-6.633874666666686);
\draw [line width=1.2pt] (5.81032533333336,-7.393874666666638)-- (6.590325333333335,-7.413874666666686);
\draw [line width=1.2pt] (6.010325333333335,-3.0138746666666862)-- (5.010325333333335,-5.013874666666686);
\draw [line width=1.2pt] (6.010325333333335,-3.0138746666666862)-- (9.010325333333338,-3.0138746666666862);
\draw [line width=1.2pt] (9.010325333333338,-3.0138746666666862)-- (8.010325333333334,-5.013874666666686);
\draw [line width=1.2pt] (8.010325333333334,-5.013874666666686)-- (5.010325333333335,-5.013874666666686);
\draw [line width=1.2pt] (8.070325333333336,-3.6672080000000156)-- (7.690325333333336,-4.427208000000016);
\draw [line width=1.2pt] (7.290325333333335,-3.647208000000016)-- (6.910325333333335,-4.407208000000016);
\draw [line width=1.2pt] (8.070325333333324,-3.667208000000041)-- (7.290325333333335,-3.647208000000016);
\draw [line width=1.2pt] (6.910325333333345,-4.407207999999996)-- (7.690325333333336,-4.427208000000016);
\draw [line width=1.2pt] (6.970325333333335,-3.6538746666666864)-- (6.590325333333335,-4.413874666666686);
\draw [line width=1.2pt] (6.190325333333336,-3.6338746666666863)-- (5.810325333333336,-4.393874666666687);
\draw [line width=1.2pt] (6.970325333333333,-3.6538746666666904)-- (6.190325333333335,-3.6338746666666877);
\draw [line width=1.2pt] (5.810325333333336,-4.393874666666687)-- (6.59032533333334,-4.413874666666678);
\draw (0.8309034874363014,-3.8488619309990755) node[anchor=north west] {$A^*$};
\draw (2.1414553452574387,-3.8488619309990755) node[anchor=north west] {$B^*$};
\draw (1.026092062005407,-6.8603427957795535) node[anchor=north west] {$A$};
\draw (2.113571263176138,-6.8603427957795535) node[anchor=north west] {$B$};
\draw [line width=1.2pt] (6,0)-- (5,-2);
\draw [line width=1.2pt] (6,0)-- (9,0);
\draw [line width=1.2pt] (9,0)-- (8,-2);
\draw [line width=1.2pt] (8,-2)-- (5,-2);
\draw [line width=1.2pt] (8.073333333333332,-0.6533333333333334)-- (7.6933333333333325,-1.4133333333333336);
\draw [line width=1.2pt] (7.293333333333333,-0.6333333333333334)-- (6.913333333333333,-1.3933333333333335);
\draw [line width=1.2pt] (8.073333333333332,-0.6533333333333342)-- (7.293333333333331,-0.6333333333333364);
\draw [line width=1.2pt] (6.913333333333333,-1.3933333333333335)-- (7.6933333333333325,-1.4133333333333336);
\draw [line width=1.2pt] (6.96,-0.64)-- (6.58,-1.4);
\draw [line width=1.2pt] (6.18,-0.62)-- (5.8,-1.38);
\draw [line width=1.2pt] (6.96,-0.64)-- (6.18,-0.62);
\draw [line width=1.2pt] (5.8,-1.38)-- (6.58,-1.4);
\draw [line width=1.2pt] (6,3)-- (5,1);
\draw [line width=1.2pt] (6,3)-- (9,3);
\draw [line width=1.2pt] (9,3)-- (8,1);
\draw [line width=1.2pt] (8,1)-- (5,1);
\draw [line width=1.2pt] (8.06,2.3466666666666662)-- (7.68,1.5866666666666667);
\draw [line width=1.2pt] (7.28,2.3666666666666667)-- (6.9,1.6066666666666667);
\draw [line width=1.2pt] (8.06,2.3466666666666307)-- (7.28,2.366666666666638);
\draw [line width=1.2pt] (6.9,1.6066666666666667)-- (7.68,1.5866666666666667);
\draw [line width=1.2pt] (6.96,2.36)-- (6.58,1.6);
\draw [line width=1.2pt] (6.18,2.38)-- (5.8,1.62);
\draw [line width=1.2pt] (6.96,2.36)-- (6.18,2.38);
\draw [line width=1.2pt] (5.8,1.62)-- (6.58,1.6);
\draw (5.850038262070444,-3.793093766836474) node[anchor=north west] {$A^*$};
\draw (7.0490537915663785,-3.820977848917775) node[anchor=north west] {$B^*$};
\draw (5.9336905083143465,-6.8603427957795535) node[anchor=north west] {$A$};
\draw (7.0490537915663785,-6.832458713698253) node[anchor=north west] {$B$};
\draw [line width=1.2pt,dotted,color=bcduew] (0.8,1.62)-- (0.9051785714285712,1);
\draw [line width=1.2pt,dotted,color=ccccff] (1.58,1.6)-- (1.6817857142857184,1);
\draw [line width=1.2pt,dotted,color=ccccff] (2.006527777777778,1)-- (1.9,1.606666666666668);
\draw [line width=1.2pt,dotted,color=ccccff] (2.7830158730158767,1)-- (2.68,1.5866666666666738);
\draw [line width=1.2pt,dotted,color=ccccff] (5.905178571428572,1)-- (5.8,1.62);
\draw [line width=1.2pt,dotted,color=ccccff] (6.6817857142857235,1)-- (6.58,1.6);
\draw [line width=1.2pt,dotted,color=ccccff] (7.006527777777778,1)-- (6.9,1.6066666666666667);
\draw [line width=1.2pt,dotted,color=ccccff] (7.7830158730158745,1)-- (7.68,1.5866666666666671);
\draw [line width=1.2pt,dotted,color=ccccff] (7.793341206349206,-5.013874666666687)-- (7.690325333333336,-4.427208000000019);
\draw [line width=1.2pt,dotted,color=ccccff] (7.016853111111115,-5.013874666666687)-- (6.910325333333341,-4.407207999999999);
\draw [line width=1.2pt,dotted,color=ccccff] (6.69211104761905,-5.013874666666687)-- (6.590325333333338,-4.41387466666668);
\draw [line width=1.2pt,dotted,color=ccccff] (5.915503904761904,-5.013874666666687)-- (5.810325333333334,-4.393874666666688);
\draw [line width=1.2pt,dotted,color=ccccff] (2.7830158730158727,-5)-- (2.68,-4.413333333333335);
\draw [line width=1.2pt,dotted,color=ccccff] (2.006527777777776,-5)-- (1.9,-4.393333333333336);
\draw [line width=1.2pt,dotted,color=ccccff] (1.6636270666666606,-5)-- (1.58,-4.4);
\draw [line width=1.2pt,dotted,color=ccccff] (0.9051785714285706,-5)-- (0.8,-4.38);
\draw [line width=1.2pt,color=ccwwff] (0.9051785714285712,1)-- (1.18,-0.62);
\draw [line width=1.2pt,color=ccwwff] (1.6817857142857184,1)-- (1.96,-0.64);
\draw [line width=1.2pt,color=ccwwff] (2.006527777777778,1)-- (2.293333333333334,-0.6333333333333366);
\draw [line width=1.2pt,color=ccwwff] (2.7830158730158767,1)-- (3.073333333333335,-0.6533333333333341);
\draw [line width=1.2pt,color=ccwwff] (2.7830158730158727,-5)-- (3.073333333333332,-6.653333333333334);
\draw [line width=1.2pt,color=ccwwff] (2.006527777777776,-5)-- (2.2933333333333317,-6.633333333333336);
\draw [line width=1.2pt,color=ccwwff] (1.6636270666666606,-5)-- (1.96,-6.64);
\draw [line width=1.2pt,color=ccwwff] (0.9051785714285706,-5)-- (1.18,-6.62);
\draw [line width=1.2pt,color=ccwwff] (5.915503904761904,-5.013874666666687)-- (6.19032533333333,-6.633874666666684);
\draw [line width=1.2pt,color=ccwwff] (6.69211104761905,-5.013874666666687)-- (6.970325333333327,-6.653874666666707);
\draw [line width=1.2pt,color=ccwwff] (7.016853111111115,-5.013874666666687)-- (7.303658666666659,-6.64720800000002);
\draw [line width=1.2pt,color=ccwwff] (7.793341206349206,-5.013874666666687)-- (8.08365866666666,-6.667208000000041);
\draw [line width=1.2pt,color=ccwwff] (7.7830158730158745,1)-- (8.073333333333332,-0.6533333333333343);
\draw [line width=1.2pt,color=ccwwff] (6.6817857142857235,1)-- (6.96,-0.64);
\draw [line width=1.2pt,color=ccwwff] (7.006527777777778,1)-- (7.29333333333333,-0.6333333333333366);
\draw [line width=1.2pt,color=ccwwff] (5.905178571428572,1)-- (6.18,-0.62);
\draw [shift={(8.197358688888876,2.6733333333333142)},line width=1.2pt,color=ffvvqq]  plot[domain=1.1727554671595342:4.314348120749328,variable=\t]({1*0.3543705977143108*cos(\t r)+0*0.3543705977143108*sin(\t r)},{0*0.3543705977143108*cos(\t r)+1*0.3543705977143108*sin(\t r)});
\draw [shift={(7.388626088888875,2.6833333333333176)},line width=1.2pt,color=ffvvqq]  plot[domain=1.2403445249793834:4.381937178569176,variable=\t]({1*0.334779636425343*cos(\t r)+0*0.334779636425343*sin(\t r)},{0*0.334779636425343*cos(\t r)+1*0.334779636425343*sin(\t r)});
\draw [line width=1.2pt,dotted,color=ffvvqq] (8.334717377777764,3)-- (8.334717377777762,1.6694347555555256);
\draw [line width=1.2pt,dotted,color=ffvvqq] (8.334717377777764,0)-- (8.334717377777762,-1.330565244444475);
\draw [line width=1.2pt,dotted,color=ffvvqq] (8.334717377777762,-3.0138746666666862)-- (8.33471737777776,-4.365090577777833);
\draw [line width=1.2pt,dotted,color=ffvvqq] (8.334717377777762,-6.013874666666685)-- (8.33471737777776,-7.365090577777833);
\draw [line width=1.2pt,color=ffvvqq] (8.334717377777762,1.6694347555555256)-- (8.334717377777764,0);
\draw [line width=1.2pt,color=ffvvqq] (8.334717377777762,-1.330565244444475)-- (8.334717377777762,-3.0138746666666862);
\draw [line width=1.2pt,color=ffvvqq] (8.33471737777776,-4.365090577777833)-- (8.334717377777762,-6.013874666666685);
\draw [shift={(7.980405177777763,-7.689482622222261)},line width=1.2pt,color=ffvvqq]  plot[domain=-2.4002501976648816:0.7413424559249115,variable=\t]({1*0.48038248674121353*cos(\t r)+0*0.48038248674121353*sin(\t r)},{0*0.48038248674121353*cos(\t r)+1*0.48038248674121353*sin(\t r)});
\draw [line width=1.2pt,dotted,color=ffvvqq] (7.647566444444432,-8.013874666666686)-- (7.703658666666672,-7.427208000000017);
\draw [line width=1.2pt,dotted,color=ffvvqq] (7.583146044444431,3)-- (7.583146044444431,1);
\draw [line width=1.2pt,dotted,color=ffvvqq] (7.583146044444431,0)-- (7.583146044444431,-2);
\draw [line width=1.2pt,dotted,color=ffvvqq] (7.583146044444432,-3.0138746666666862)-- (7.583146044444431,-5.013874666666688);
\draw [line width=1.2pt,color=ffvvqq] (7.583146044444431,1)-- (7.583146044444431,0);
\draw [line width=1.2pt,color=ffvvqq] (7.583146044444431,-2)-- (7.583146044444432,-3.0138746666666862);
\draw [line width=1.2pt,color=ffvvqq] (7.583146044444431,-5.013874666666688)-- (7.583146044444431,-6.013874666666684);
\draw [line width=1.2pt,dotted,color=ffvvqq] (7.583146044444432,-6.013874666666685)-- (7.583146044444431,-8.013874666666688);
\draw [line width=1.2pt,dotted,color=ffvvqq] (6.917468577777766,-8.013874666666686)-- (6.923658666666696,-7.407207999999949);
\draw [shift={(7.250307311111099,-8.013874666666686)},line width=1.2pt,color=ffvvqq]  plot[domain=3.141592653589793:6.283185307179586,variable=\t]({1*0.33283873333333336*cos(\t r)+0*0.33283873333333336*sin(\t r)},{0*0.33283873333333336*cos(\t r)+1*0.33283873333333336*sin(\t r)});
\draw [line width=1.2pt] (0,3.6336251407407376)-- (9.601651911111096,3.590678207407404);
\draw [line width=1.2pt] (4.383599511111101,4.771718874074067)-- (4.426546444444434,-8.541830459259225);
\draw (1.4443532932249188,4.572130857553743) node[anchor=north west] {$|\rho_{AB}^2\rangle\langle\rho_{AB}^2|$};
\draw (5.013515799631421,4.6278990217163445) node[anchor=north west] {$Tr_{\mathcal{H}_B\otimes\mathcal{H}^*_B}|\rho_{AB}^2\rangle\langle\rho_{AB}^2|$};
\end{tikzpicture}

\caption{Construction of replica manifold of $\ket{\rho_{AB}^{m/2}}\bra{\rho_{AB}^{m/2}}$ and its trace for $m=4$. The corresponding tracing out can be found by the colored lines. }
\label{fig204}
\end{figure}
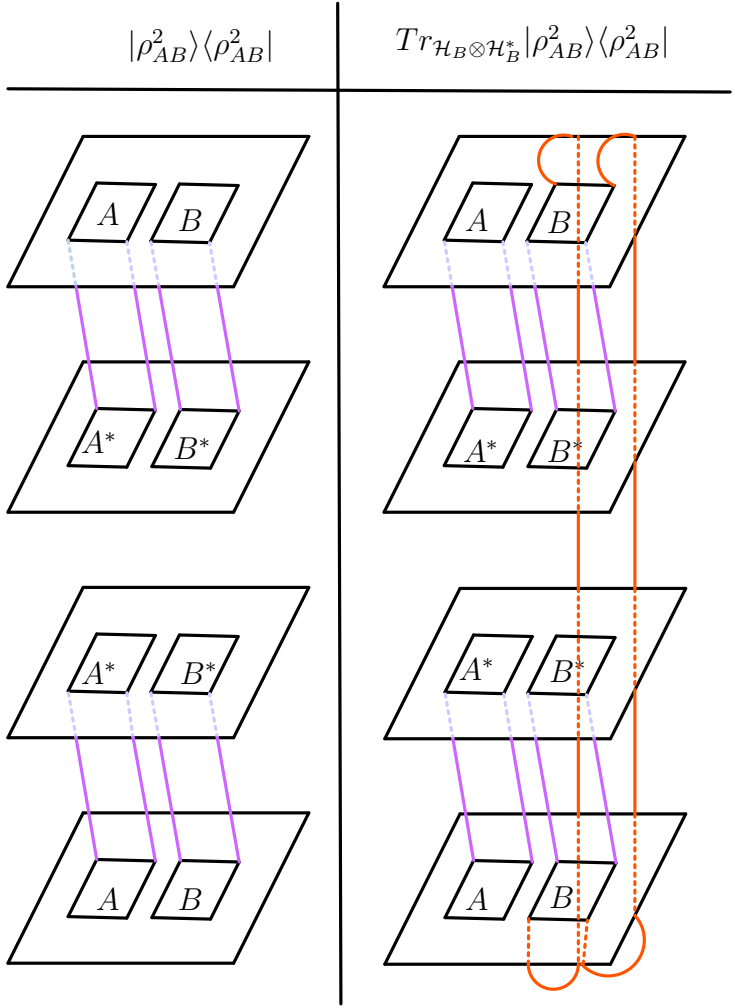

In Fig.\ref{fig204} the $\ket{\rho_{AB}^{m/2}}\bra{\rho_{AB}^{m/2}}$ and its trace on $\mathcal{H}_B\otimes\mathcal{H}_{B^*}$ are depicted for $m=4$. The positions of $(\mathcal{H}_A,\mathcal{H}_B)$ and $(\mathcal{H}_A^*,\mathcal{H}_B^*)$ in $\ket{\rho_{AB}}$ replica manifold are different in the $\bra{\rho_{AB}}$ hermitian replica manifold that can be seen in Fig.\ref{fig204}.

\begin{figure}[!h]
\centering
\captionsetup{width=.8\linewidth}

\definecolor{qqqqff}{rgb}{0,0,1}
\definecolor{ffffff}{rgb}{1,1,1}
\definecolor{ffvvqq}{rgb}{1,0.3333333333333333,0}
\definecolor{ccwwff}{rgb}{0.8,0.4,1}
\definecolor{ccccff}{rgb}{0.8,0.8,1}
\definecolor{bcduew}{rgb}{0.7372549019607844,0.8313725490196079,0.9019607843137255}
\begin{tikzpicture}[line cap=round,line join=round,>=triangle 45,x=1cm,y=1cm]
\clip(-3,-8.8) rectangle (9.3,5.4);
\fill[line width=1.2pt,color=ffffff] (6.179577413940357,2.7419211063992823) -- (6.956951823479654,2.730654810608858) -- (6.956951823479654,2.370133345315273) -- (6.179577413940357,2.3476007537344237) -- cycle;
\fill[line width=1.2pt,color=ffffff] (5.7908902091706995,-1.3928094486865268) -- (6.568264618709994,-1.404075744476951) -- (6.568264618709994,-1.7645972097705367) -- (5.7908902091706995,-1.7871298013513859) -- cycle;
\fill[line width=1.2pt,color=ffffff] (6.185210561835559,-3.2658311238446074) -- (6.962584971374854,-3.2770974196350315) -- (6.962584971374854,-3.6376188849286173) -- (6.185210561835559,-3.6601514765094665) -- cycle;
\fill[line width=1.2pt,color=ffffff] (5.819055948646757,-7.392111957087596) -- (6.596430358186053,-7.40337825287802) -- (6.596430358186053,-7.763899718171606) -- (5.819055948646757,-7.786432309752455) -- cycle;
\fill[line width=1.2pt,color=ffffff] (1.1857918048346445,2.761637124032522) -- (1.9631662143739401,2.7503708282420978) -- (1.9631662143739401,2.389849362948512) -- (1.1857918048346445,2.367316771367663) -- cycle;
\fill[line width=1.2pt,color=ffffff] (0.7914714521697852,-1.3928094486865275) -- (1.5688458617090797,-1.4040757444769516) -- (1.5688458617090797,-1.7645972097705371) -- (0.7914714521697852,-1.7871298013513863) -- cycle;
\fill[line width=1.2pt,color=ffffff] (1.1810975149219678,-3.2564425440192535) -- (1.9584719244612623,-3.267708839809677) -- (1.9584719244612623,-3.6282303051032634) -- (1.1810975149219678,-3.650762896684112) -- cycle;
\fill[line width=1.2pt,color=ffffff] (0.7961657420824602,-7.39211195708759) -- (1.573540151621755,-7.4033782528780145) -- (1.573540151621755,-7.763899718171601) -- (0.7961657420824602,-7.786432309752449) -- cycle;
\draw [line width=1.2pt] (1,0)-- (0,-2);
\draw [line width=1.2pt] (1,0)-- (4,0);
\draw [line width=1.2pt] (4,0)-- (3,-2);
\draw [line width=1.2pt] (3,-2)-- (0,-2);
\draw [line width=1.2pt] (3.0733333333333355,-0.6533333333333333)-- (2.6933333333333356,-1.413333333333333);
\draw [line width=1.2pt] (2.2933333333333352,-0.6333333333333337)-- (1.9133333333333353,-1.3933333333333335);
\draw [line width=1.2pt] (3.073333333333335,-0.6533333333333342)-- (2.293333333333334,-0.6333333333333367);
\draw [line width=1.2pt] (1.9133333333333353,-1.3933333333333335)-- (2.6933333333333356,-1.413333333333333);
\draw [line width=1.2pt] (1.96,-0.64)-- (1.58,-1.4);
\draw [line width=1.2pt] (1.18,-0.62)-- (0.8,-1.38);
\draw [line width=1.2pt] (1.96,-0.64)-- (1.18,-0.62);
\draw [line width=1.2pt] (0.8,-1.38)-- (1.58,-1.4);
\draw [line width=1.2pt] (1,3)-- (0,1);
\draw [line width=1.2pt] (1,3)-- (4,3);
\draw [line width=1.2pt] (4,3)-- (3,1);
\draw [line width=1.2pt] (3,1)-- (0,1);
\draw [line width=1.2pt] (3.06,2.3466666666666667)-- (2.68,1.5866666666666671);
\draw [line width=1.2pt] (2.28,2.366666666666667)-- (1.9,1.606666666666667);
\draw [line width=1.2pt] (3.06,2.3466666666666582)-- (2.28,2.3666666666666565);
\draw [line width=1.2pt] (1.9,1.606666666666668)-- (2.68,1.5866666666666804);
\draw [line width=1.2pt] (1.96,2.36)-- (1.58,1.6);
\draw [line width=1.2pt] (1.18,2.38)-- (0.8,1.62);
\draw [line width=1.2pt] (1.96,2.36)-- (1.18,2.38);
\draw [line width=1.2pt] (0.8,1.62)-- (1.58,1.6);
\draw (5.931483657691463,2.1249586250583037) node[anchor=north west] {$A$};
\draw (6.996456195754374,2.1830480362253715) node[anchor=north west] {$B$};
\draw (5.892757383580085,-0.8763276185735236) node[anchor=north west] {$A^*$};
\draw (7.054545606921441,-0.8763276185735236) node[anchor=north west] {$B^*$};
\draw [line width=1.2pt] (1,-6)-- (0,-8);
\draw [line width=1.2pt] (1,-6)-- (4,-6);
\draw [line width=1.2pt] (4,-6)-- (3,-8);
\draw [line width=1.2pt] (3,-8)-- (0,-8);
\draw [line width=1.2pt] (3.0733333333333355,-6.653333333333333)-- (2.693333333333335,-7.413333333333334);
\draw [line width=1.2pt] (2.2933333333333352,-6.633333333333334)-- (1.9133333333333353,-7.3933333333333335);
\draw [line width=1.2pt] (3.0733333333333355,-6.653333333333333)-- (2.2933333333333352,-6.633333333333334);
\draw [line width=1.2pt] (1.9133333333333544,-7.393333333333295)-- (2.6933333333333542,-7.413333333333296);
\draw [line width=1.2pt] (1.96,-6.64)-- (1.58,-7.4);
\draw [line width=1.2pt] (1.18,-6.62)-- (0.8,-7.38);
\draw [line width=1.2pt] (1.96,-6.64)-- (1.18,-6.62);
\draw [line width=1.2pt] (0.8,-7.38)-- (1.58,-7.4);
\draw [line width=1.2pt] (1,-3)-- (0,-5);
\draw [line width=1.2pt] (1,-3)-- (4,-3);
\draw [line width=1.2pt] (4,-3)-- (3,-5);
\draw [line width=1.2pt] (3,-5)-- (0,-5);
\draw [line width=1.2pt] (3.06,-3.6533333333333338)-- (2.68,-4.413333333333333);
\draw [line width=1.2pt] (2.28,-3.6333333333333333)-- (1.9,-4.3933333333333335);
\draw [line width=1.2pt] (3.06,-3.6533333333333458)-- (2.28,-3.633333333333346);
\draw [line width=1.2pt] (1.9,-4.3933333333333335)-- (2.68,-4.413333333333333);
\draw [line width=1.2pt] (1.96,-3.64)-- (1.58,-4.4);
\draw [line width=1.2pt] (1.18,-3.62)-- (0.8,-4.38);
\draw [line width=1.2pt] (1.96,-3.64)-- (1.18,-3.62);
\draw [line width=1.2pt] (0.8,-4.38)-- (1.58,-4.4);
\draw (1.0326099826020738,2.163684899169682) node[anchor=north west] {$A$};
\draw (2.116945657720674,2.1249586250583037) node[anchor=north west] {$B$};
\draw (0.838978612045181,-0.8182382074064559) node[anchor=north west] {$A^*$};
\draw (2.0782193836092953,-0.8376013444621451) node[anchor=north west] {$B^*$};
\draw [line width=1.2pt] (6.010325333333335,-6.013874666666686)-- (5.010325333333335,-8.013874666666686);
\draw [line width=1.2pt] (6.010325333333335,-6.013874666666686)-- (9.010325333333338,-6.013874666666686);
\draw [line width=1.2pt] (9.010325333333338,-6.013874666666686)-- (8.010325333333334,-8.013874666666686);
\draw [line width=1.2pt] (8.010325333333334,-8.013874666666686)-- (5.010325333333335,-8.013874666666686);
\draw [line width=1.2pt] (8.083658666666668,-6.667208000000016)-- (7.703658666666668,-7.427208000000015);
\draw [line width=1.2pt] (7.303658666666669,-6.647208000000016)-- (6.923658666666669,-7.407208000000016);
\draw [line width=1.2pt] (8.083658666666647,-6.66720800000006)-- (7.303658666666669,-6.647208000000016);
\draw [line width=1.2pt] (6.923658666666745,-7.407207999999866)-- (7.703658666666668,-7.427208000000015);
\draw [line width=1.2pt] (6.970325333333335,-6.653874666666686)-- (6.590325333333335,-7.413874666666686);
\draw [line width=1.2pt] (6.190325333333336,-6.633874666666686)-- (5.810325333333336,-7.393874666666687);
\draw [line width=1.2pt] (6.970325333333314,-6.653874666666729)-- (6.190325333333336,-6.633874666666686);
\draw [line width=1.2pt] (5.810325333333391,-7.3938746666665764)-- (6.590325333333335,-7.413874666666686);
\draw [line width=1.2pt] (6.010325333333335,-3.0138746666666862)-- (5.010325333333335,-5.013874666666686);
\draw [line width=1.2pt] (6.010325333333335,-3.0138746666666862)-- (9.010325333333338,-3.0138746666666862);
\draw [line width=1.2pt] (9.010325333333338,-3.0138746666666862)-- (8.010325333333334,-5.013874666666686);
\draw [line width=1.2pt] (8.010325333333334,-5.013874666666686)-- (5.010325333333335,-5.013874666666686);
\draw [line width=1.2pt] (8.070325333333336,-3.6672080000000156)-- (7.690325333333336,-4.427208000000016);
\draw [line width=1.2pt] (7.290325333333335,-3.647208000000016)-- (6.910325333333335,-4.407208000000016);
\draw [line width=1.2pt] (8.070325333333315,-3.6672080000000573)-- (7.290325333333335,-3.647208000000016);
\draw [line width=1.2pt] (6.9103253333333585,-4.40720799999997)-- (7.690325333333336,-4.427208000000016);
\draw [line width=1.2pt] (6.970325333333335,-3.6538746666666864)-- (6.590325333333335,-4.413874666666686);
\draw [line width=1.2pt] (6.190325333333336,-3.6338746666666863)-- (5.810325333333336,-4.393874666666687);
\draw [line width=1.2pt] (6.970325333333333,-3.6538746666666904)-- (6.190325333333335,-3.6338746666666877);
\draw [line width=1.2pt] (5.810325333333336,-4.393874666666687)-- (6.59032533333334,-4.413874666666678);
\draw (0.8583417491008702,-3.9163401363167294) node[anchor=north west] {$A^*$};
\draw (2.0975825206649845,-3.877613862205351) node[anchor=north west] {$B^*$};
\draw (1.051973119657763,-6.801447557614422) node[anchor=north west] {$A$};
\draw (2.116945657720674,-6.878900105837179) node[anchor=north west] {$B$};
\draw [line width=1.2pt] (6,0)-- (5,-2);
\draw [line width=1.2pt] (6,0)-- (9,0);
\draw [line width=1.2pt] (9,0)-- (8,-2);
\draw [line width=1.2pt] (8,-2)-- (5,-2);
\draw [line width=1.2pt] (8.073333333333332,-0.6533333333333334)-- (7.6933333333333325,-1.4133333333333336);
\draw [line width=1.2pt] (7.293333333333333,-0.6333333333333334)-- (6.913333333333333,-1.3933333333333335);
\draw [line width=1.2pt] (8.073333333333332,-0.6533333333333342)-- (7.293333333333331,-0.6333333333333364);
\draw [line width=1.2pt] (6.913333333333333,-1.3933333333333335)-- (7.6933333333333325,-1.4133333333333336);
\draw [line width=1.2pt] (6.96,-0.64)-- (6.58,-1.4);
\draw [line width=1.2pt] (6.18,-0.62)-- (5.8,-1.38);
\draw [line width=1.2pt] (6.96,-0.64)-- (6.18,-0.62);
\draw [line width=1.2pt] (5.8,-1.38)-- (6.58,-1.4);
\draw [line width=1.2pt] (6,3)-- (5,1);
\draw [line width=1.2pt] (6,3)-- (9,3);
\draw [line width=1.2pt] (9,3)-- (8,1);
\draw [line width=1.2pt] (8,1)-- (5,1);
\draw [line width=1.2pt] (8.06,2.3466666666666662)-- (7.68,1.5866666666666667);
\draw [line width=1.2pt] (7.28,2.3666666666666667)-- (6.9,1.6066666666666667);
\draw [line width=1.2pt] (8.06,2.3466666666666036)-- (7.28,2.3666666666666183);
\draw [line width=1.2pt] (6.9,1.6066666666666667)-- (7.68,1.5866666666666667);
\draw [line width=1.2pt] (6.96,2.36)-- (6.58,1.6);
\draw [line width=1.2pt] (6.18,2.38)-- (5.8,1.62);
\draw [line width=2.8pt] (6.96,2.36)-- (6.18,2.38);
\draw [line width=1.2pt] (5.8,1.62)-- (6.58,1.6);
\draw (5.815304835357328,-3.8582507251496616) node[anchor=north west] {$A^*$};
\draw (7.073908743977131,-3.877613862205351) node[anchor=north west] {$B^*$};
\draw (5.912120520635773,-6.878900105837179) node[anchor=north west] {$A$};
\draw (7.073908743977131,-6.859536968781489) node[anchor=north west] {$B$};
\draw [line width=1.2pt,dotted,color=bcduew] (0.8,1.62)-- (0.9051785714285712,1);
\draw [line width=1.2pt,dotted,color=ccccff] (1.58,1.6)-- (1.6817857142857184,1);
\draw [line width=1.2pt,dotted,color=ccccff] (2.006527777777778,1)-- (1.9,1.6066666666666676);
\draw [line width=1.2pt,dotted,color=ccccff] (2.7830158730158803,1)-- (2.68,1.58666666666668);
\draw [line width=1.2pt,dotted,color=ccccff] (5.9051785714285705,1)-- (5.8,1.62);
\draw [line width=1.2pt,dotted,color=ccccff] (6.681785714285729,1)-- (6.58,1.6);
\draw [line width=1.2pt,dotted,color=ccccff] (7.006527777777778,1)-- (6.9,1.6066666666666667);
\draw [line width=1.2pt,dotted,color=ccccff] (7.7830158730158745,1)-- (7.68,1.5866666666666671);
\draw [line width=1.2pt,dotted,color=ccccff] (7.793341206349204,-5.013874666666687)-- (7.690325333333337,-4.427208000000022);
\draw [line width=1.2pt,dotted,color=ccccff] (7.016853111111129,-5.013874666666687)-- (6.910325333333354,-4.407207999999974);
\draw [line width=1.2pt,dotted,color=ccccff] (6.692111047619045,-5.013874666666687)-- (6.590325333333338,-4.41387466666668);
\draw [line width=1.2pt,dotted,color=ccccff] (5.915503904761905,-5.013874666666687)-- (5.810325333333334,-4.393874666666688);
\draw [line width=1.2pt,dotted,color=ccccff] (2.7830158730158727,-5)-- (2.68,-4.413333333333335);
\draw [line width=1.2pt,dotted,color=ccccff] (2.006527777777776,-5)-- (1.9,-4.393333333333336);
\draw [line width=1.2pt,dotted,color=ccccff] (1.6636270666666606,-5)-- (1.58,-4.4);
\draw [line width=1.2pt,dotted,color=ccccff] (0.9051785714285706,-5)-- (0.8,-4.38);
\draw [line width=1.2pt,color=ccwwff] (0.9051785714285712,1)-- (1.18,-0.62);
\draw [line width=1.2pt,color=ccwwff] (1.6817857142857184,1)-- (1.96,-0.64);
\draw [line width=1.2pt,color=ccwwff] (2.006527777777778,1)-- (2.293333333333334,-0.6333333333333366);
\draw [line width=1.2pt,color=ccwwff] (2.7830158730158803,1)-- (3.073333333333335,-0.6533333333333341);
\draw [line width=1.2pt,color=ccwwff] (2.7830158730158727,-5)-- (3.073333333333332,-6.653333333333334);
\draw [line width=1.2pt,color=ccwwff] (2.006527777777776,-5)-- (2.2933333333333317,-6.633333333333336);
\draw [line width=1.2pt,color=ccwwff] (1.6636270666666606,-5)-- (1.96,-6.64);
\draw [line width=1.2pt,color=ccwwff] (0.9051785714285706,-5)-- (1.18,-6.62);
\draw [line width=1.2pt,color=ccwwff] (5.915503904761905,-5.013874666666687)-- (6.19032533333333,-6.6338746666666815);
\draw [line width=1.2pt,color=ccwwff] (6.692111047619045,-5.013874666666687)-- (6.970325333333318,-6.653874666666725);
\draw [line width=1.2pt,color=ccwwff] (7.016853111111129,-5.013874666666687)-- (7.30365866666666,-6.647208000000019);
\draw [line width=1.2pt,color=ccwwff] (7.793341206349204,-5.013874666666687)-- (8.083658666666649,-6.667208000000062);
\draw [line width=1.2pt,color=ccwwff] (7.7830158730158745,1)-- (8.073333333333332,-0.6533333333333343);
\draw [line width=1.2pt,color=ccwwff] (6.681785714285729,1)-- (6.96,-0.64);
\draw [line width=1.2pt,color=ccwwff] (7.006527777777778,1)-- (7.29333333333333,-0.6333333333333366);
\draw [line width=1.2pt,color=ccwwff] (5.9051785714285705,1)-- (6.18,-0.62);
\draw [shift={(8.197358688888867,2.6733333333333)},line width=1.2pt,color=ffvvqq]  plot[domain=1.1727554671595288:4.314348120749322,variable=\t]({1*0.3543705977143274*cos(\t r)+0*0.3543705977143274*sin(\t r)},{0*0.3543705977143274*cos(\t r)+1*0.3543705977143274*sin(\t r)});
\draw [shift={(7.388626088888871,2.683333333333308)},line width=1.2pt,color=ffvvqq]  plot[domain=1.2403445249793794:4.3819371785691725,variable=\t]({1*0.33477963642535363*cos(\t r)+0*0.33477963642535363*sin(\t r)},{0*0.33477963642535363*cos(\t r)+1*0.33477963642535363*sin(\t r)});
\draw [line width=1.2pt,dotted,color=ffvvqq] (8.334717377777764,3)-- (8.334717377777762,1.6694347555555256);
\draw [line width=1.2pt,dotted,color=ffvvqq] (8.334717377777764,0)-- (8.334717377777762,-1.330565244444475);
\draw [line width=1.2pt,dotted,color=ffvvqq] (8.334717377777762,-3.0138746666666862)-- (8.33471737777776,-4.365090577777833);
\draw [line width=1.2pt,dotted,color=ffvvqq] (8.334717377777762,-6.013874666666685)-- (8.33471737777776,-7.365090577777833);
\draw [line width=1.2pt,color=ffvvqq] (8.334717377777762,1.6694347555555256)-- (8.334717377777764,0);
\draw [line width=1.2pt,color=ffvvqq] (8.334717377777762,-1.330565244444475)-- (8.334717377777762,-3.0138746666666862);
\draw [line width=1.2pt,color=ffvvqq] (8.33471737777776,-4.365090577777833)-- (8.334717377777762,-6.013874666666685);
\draw [shift={(7.980405177777763,-7.689482622222261)},line width=1.2pt,color=ffvvqq]  plot[domain=-2.4002501976648816:0.7413424559249115,variable=\t]({1*0.48038248674121353*cos(\t r)+0*0.48038248674121353*sin(\t r)},{0*0.48038248674121353*cos(\t r)+1*0.48038248674121353*sin(\t r)});
\draw [line width=1.2pt,dotted,color=ffvvqq] (7.647566444444432,-8.013874666666686)-- (7.70365866666667,-7.427208000000019);
\draw [line width=1.2pt,dotted,color=ffvvqq] (7.583146044444431,3)-- (7.583146044444431,1);
\draw [line width=1.2pt,dotted,color=ffvvqq] (7.583146044444431,0)-- (7.583146044444431,-2);
\draw [line width=1.2pt,dotted,color=ffvvqq] (7.583146044444432,-3.0138746666666862)-- (7.583146044444431,-5.013874666666688);
\draw [line width=1.2pt,color=ffvvqq] (7.583146044444431,1)-- (7.583146044444431,0);
\draw [line width=1.2pt,color=ffvvqq] (7.583146044444431,-2)-- (7.583146044444432,-3.0138746666666862);
\draw [line width=1.2pt,color=ffvvqq] (7.583146044444431,-5.013874666666688)-- (7.583146044444431,-6.013874666666686);
\draw [line width=1.2pt,dotted,color=ffvvqq] (7.583146044444433,-6.013874666666685)-- (7.583146044444431,-8.013874666666688);
\draw [shift={(7.250307311111099,-8.013874666666686)},line width=1.2pt,color=ffvvqq]  plot[domain=3.141592653589793:6.283185307179586,variable=\t]({1*0.33283873333333336*cos(\t r)+0*0.33283873333333336*sin(\t r)},{0*0.33283873333333336*cos(\t r)+1*0.33283873333333336*sin(\t r)});
\draw [line width=1.2pt,dotted,color=ffvvqq] (6.917468577777766,-8.013874666666686)-- (6.923658666666734,-7.407207999999869);
\draw [shift={(3.144467273214082,2.6733333333333293)},line width=1.2pt,color=ffvvqq]  plot[domain=1.317765107597007:4.4593577611868,variable=\t]({1*0.3374104790241954*cos(\t r)+0*0.3374104790241954*sin(\t r)},{0*0.3374104790241954*cos(\t r)+1*0.3374104790241954*sin(\t r)});
\draw [shift={(2.379581280787704,2.6833333333333282)},line width=1.2pt,color=ffvvqq]  plot[domain=1.2661202790276813:4.407712932617475,variable=\t]({1*0.33195513139745514*cos(\t r)+0*0.33195513139745514*sin(\t r)},{0*0.33195513139745514*cos(\t r)+1*0.33195513139745514*sin(\t r)});
\draw [line width=1.2pt,dotted,color=ffvvqq] (3.2289345464281696,3)-- (3.2289345464281696,1.4578690928563394);
\draw [line width=1.2pt,color=ffvvqq] (3.2289345464281696,1.4578690928563394)-- (3.2289345464281696,0);
\draw [line width=1.2pt,color=ffvvqq] (3.2289345464281696,-1.5421309071436606)-- (3.2289345464281696,-3);
\draw [line width=1.2pt,color=ffvvqq] (3.2289345464281696,-4.542130907143661)-- (3.2289345464281696,-6);
\draw [line width=1.2pt,dotted,color=ffvvqq] (3.2289345464281696,0)-- (3.2289345464281696,-1.5421309071436606);
\draw [line width=1.2pt,dotted,color=ffvvqq] (3.2289345464281696,-3)-- (3.2289345464281696,-4.542130907143661);
\draw [line width=1.2pt,dotted,color=ffvvqq] (3.2289345464281696,-6)-- (3.2289345464281696,-7.542130907143661);
\draw [shift={(2.953191956957528,-7.771065453571829)},line width=1.2pt,color=ffvvqq]  plot[domain=-2.448678396384478:0.6929142572053153,variable=\t]({1*0.35839224628365945*cos(\t r)+0*0.35839224628365945*sin(\t r)},{0*0.35839224628365945*cos(\t r)+1*0.35839224628365945*sin(\t r)});
\draw [line width=1.2pt,dotted,color=ffvvqq] (2.677449367486886,-8)-- (2.6933333333333516,-7.413333333333299);
\draw [line width=1.2pt,dotted,color=ffvvqq] (2.479162561575414,3)-- (2.4791625615754134,1);
\draw [line width=1.2pt,dotted,color=ffvvqq] (2.4791625615754134,0)-- (2.4791625615754134,-2);
\draw [line width=1.2pt,dotted,color=ffvvqq] (2.4791625615754134,-3)-- (2.4791625615754134,-5);
\draw [line width=1.2pt,dotted,color=ffvvqq] (2.4791625615754134,-6)-- (2.4791625615754134,-8);
\draw [line width=1.2pt,color=ffvvqq] (2.4791625615754134,-5)-- (2.4791625615754134,-6);
\draw [line width=1.2pt,color=ffvvqq] (2.4791625615754134,-2)-- (2.4791625615754134,-3);
\draw [line width=1.2pt,color=ffvvqq] (2.4791625615754134,1)-- (2.4791625615754134,0);
\draw [shift={(2.1817323527082055,-8)},line width=1.2pt,color=ffvvqq]  plot[domain=3.141592653589793:6.283185307179586,variable=\t]({1*0.29743020886720783*cos(\t r)+0*0.29743020886720783*sin(\t r)},{0*0.29743020886720783*cos(\t r)+1*0.29743020886720783*sin(\t r)});
\draw [line width=1.2pt,dotted,color=ffvvqq] (1.8843021438409977,-8)-- (1.913333333333351,-7.393333333333298);
\draw [line width=1.2pt,color=qqqqff] (6.179577413940357,2.7419211063992823)-- (6.956951823479654,2.730654810608858);
\draw [line width=1.2pt,color=qqqqff] (6.956951823479654,2.730654810608858)-- (6.956951823479654,2.370133345315273);
\draw [line width=1.2pt,color=qqqqff] (6.956951823479654,2.370133345315273)-- (6.179577413940357,2.3476007537344237);
\draw [line width=1.2pt,color=qqqqff] (6.179577413940357,2.3476007537344237)-- (6.179577413940357,2.7419211063992823);
\draw [line width=1.2pt,color=qqqqff] (5.7908902091706995,-1.3928094486865268)-- (6.568264618709994,-1.404075744476951);
\draw [line width=1.2pt,color=qqqqff] (6.568264618709994,-1.404075744476951)-- (6.568264618709994,-1.7645972097705367);
\draw [line width=1.2pt,color=qqqqff] (6.568264618709994,-1.7645972097705367)-- (5.7908902091706995,-1.7871298013513859);
\draw [line width=1.2pt,color=qqqqff] (5.7908902091706995,-1.7871298013513859)-- (5.7908902091706995,-1.3928094486865268);
\draw [line width=1.2pt,color=qqqqff] (6.185210561835559,-3.2658311238446074)-- (6.962584971374854,-3.2770974196350315);
\draw [line width=1.2pt,color=qqqqff] (6.962584971374854,-3.2770974196350315)-- (6.962584971374854,-3.6376188849286173);
\draw [line width=1.2pt,color=qqqqff] (6.962584971374854,-3.6376188849286173)-- (6.185210561835559,-3.6601514765094665);
\draw [line width=1.2pt,color=qqqqff] (6.185210561835559,-3.6601514765094665)-- (6.185210561835559,-3.2658311238446074);
\draw [line width=1.2pt,color=qqqqff] (5.819055948646757,-7.392111957087596)-- (6.596430358186053,-7.40337825287802);
\draw [line width=1.2pt,color=qqqqff] (6.596430358186053,-7.40337825287802)-- (6.596430358186053,-7.763899718171606);
\draw [line width=1.2pt,color=qqqqff] (6.596430358186053,-7.763899718171606)-- (5.819055948646757,-7.786432309752455);
\draw [line width=1.2pt,color=qqqqff] (5.819055948646757,-7.786432309752455)-- (5.819055948646757,-7.392111957087596);
\draw [line width=1.2pt,color=qqqqff] (1.1857918048346445,2.761637124032522)-- (1.9631662143739401,2.7503708282420978);
\draw [line width=1.2pt,color=qqqqff] (1.9631662143739401,2.7503708282420978)-- (1.9631662143739401,2.389849362948512);
\draw [line width=1.2pt,color=qqqqff] (1.9631662143739401,2.389849362948512)-- (1.1857918048346445,2.367316771367663);
\draw [line width=1.2pt,color=qqqqff] (1.1857918048346445,2.367316771367663)-- (1.1857918048346445,2.761637124032522);
\draw [line width=1.2pt,color=qqqqff] (0.7914714521697852,-1.3928094486865275)-- (1.5688458617090797,-1.4040757444769516);
\draw [line width=1.2pt,color=qqqqff] (1.5688458617090797,-1.4040757444769516)-- (1.5688458617090797,-1.7645972097705371);
\draw [line width=1.2pt,color=qqqqff] (1.5688458617090797,-1.7645972097705371)-- (0.7914714521697852,-1.7871298013513863);
\draw [line width=1.2pt,color=qqqqff] (0.7914714521697852,-1.7871298013513863)-- (0.7914714521697852,-1.3928094486865275);
\draw [line width=1.2pt,color=qqqqff] (1.1810975149219678,-3.2564425440192535)-- (1.9584719244612623,-3.267708839809677);
\draw [line width=1.2pt,color=qqqqff] (1.9584719244612623,-3.267708839809677)-- (1.9584719244612623,-3.6282303051032634);
\draw [line width=1.2pt,color=qqqqff] (1.9584719244612623,-3.6282303051032634)-- (1.1810975149219678,-3.650762896684112);
\draw [line width=1.2pt,color=qqqqff] (1.1810975149219678,-3.650762896684112)-- (1.1810975149219678,-3.2564425440192535);
\draw [line width=1.2pt,color=qqqqff] (0.7961657420824602,-7.39211195708759)-- (1.573540151621755,-7.4033782528780145);
\draw [line width=1.2pt,color=qqqqff] (1.573540151621755,-7.4033782528780145)-- (1.573540151621755,-7.763899718171601);
\draw [line width=1.2pt,color=qqqqff] (1.573540151621755,-7.763899718171601)-- (0.7961657420824602,-7.786432309752449);
\draw [line width=1.2pt,color=qqqqff] (0.7961657420824602,-7.786432309752449)-- (0.7961657420824602,-7.39211195708759);
\draw (1.3230570384374132,2.802668422007426) node[anchor=north west] {$1$};
\draw (5.970209931802842,-7.382341669285098) node[anchor=north west] {$1$};
\draw (1.3230570384374132,-3.2192672023119178) node[anchor=north west] {$3$};
\draw (6.00893620591422,-1.3604060449657538) node[anchor=north west] {$3$};
\draw (1.0132468455463846,-1.3604060449657538) node[anchor=north west] {$2$};
\draw (6.357472672916628,-3.2192672023119178) node[anchor=north west] {$2$};
\draw (1.0132468455463846,-7.343615395173719) node[anchor=north west] {$4$};
\draw (6.318746398805248,2.8220315590631153) node[anchor=north west] {$4$};
\draw [->,line width=1.2pt] (-2,4) -- (0,4);
\draw [->,line width=1.2pt] (-2,4) -- (-2,2);
\draw (-1.1747876417465049,3.9450935082930894) node[anchor=north west] {$n$};
\draw (-2.6851123320902692,3.151204889009832) node[anchor=north west] {$m$};
\end{tikzpicture}

\caption{The replica manifold $Z_{n,m}$ with $n=2$ and $m=4$.  Tracing out on $\mathcal{H}_A\otimes\mathcal{H}_{A^*}$ has been done by gluing the different parts of the replicas by the same number.}
\label{fig205}
\end{figure}
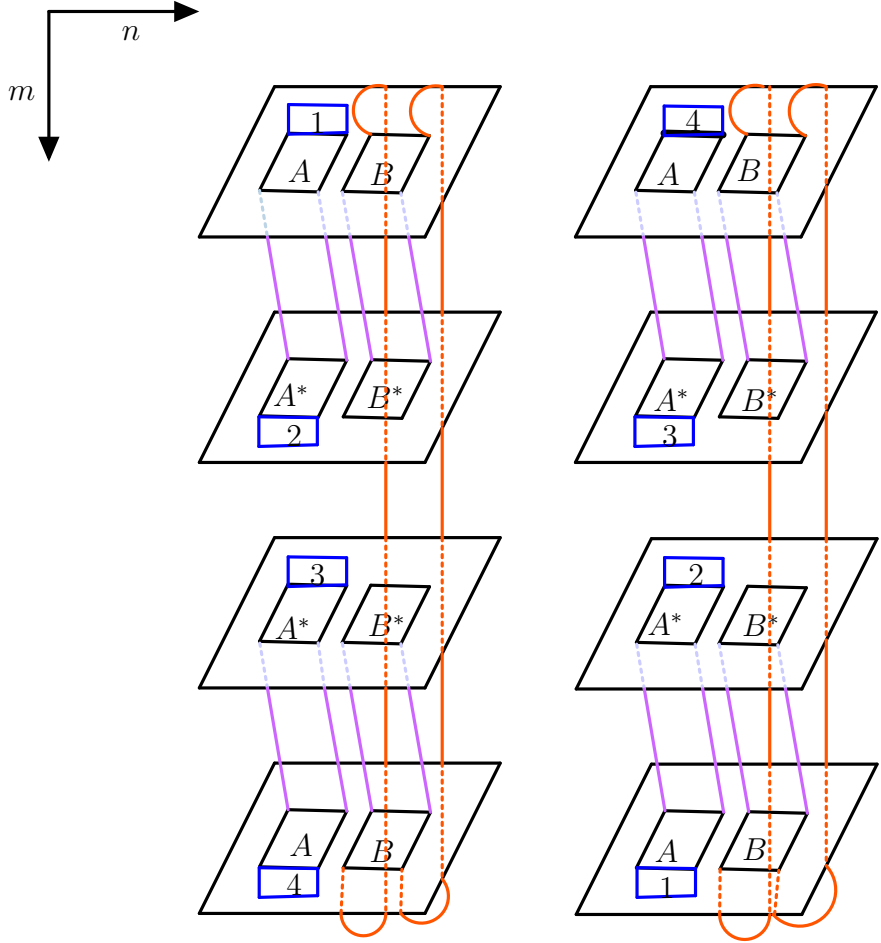

The tracing out on $\mathcal{H}_B\otimes\mathcal{H}_{B^*}$ corresponds to glue intervals $B$ (and $B^*$) on the sheets in the $\ket{\rho_{AB}}$ and $\bra{\rho_{AB}}$ in Fig.\ref{fig204}. The effect of $n$ on the construction of the $Z_{n,m}$ manifold is shown in Fig.\ref{fig205}.

For $n=2$, the construction of $\mathrm{Tr}_{BB^*}\ket{\rho_{AB}^{m/2}}\bra{\rho_{AB}^{m/2}}$ can be doubled in $Z_{2,4}$. Tracing out on $\mathcal{H}_A\otimes\mathcal{H}_{A^*}$ has been done by gluing the different parts of the replicas by the same number.

The authors in \cite{CFT01,CFT02} expressed the Renyi entropy in 2d CFT by the correlation functions of the twist operators. The idea in \cite{CFT01,CFT02} can be used to express the Renyi entropy (\ref{Rnm04}) using the correlation functions of the twist operators as follows
\begin{equation}\label{Rnm05}
 S_n(AA^*)_{\psi_m}=\frac{1}{1-n}\log\frac{\big\langle\sigma_{gA}(x_1)\sigma_{{gA}^{-1}}(x_2)
 \sigma_{gB}(x_3)\sigma_{{gB}^{-1}}(x_4)\big\rangle_{CFT^{\otimes mn}}}
 {\big(\big\langle\sigma_{gm}(x_1)\sigma_{{gm}^{-1}}(x_2)
 \sigma_{gm}(x_3)\sigma_{{gm}^{-1}}(x_4)    \big\rangle_{CFT^{\otimes m}}\big)^n }
\end{equation}
where $x_i$s are the endpoints of the intervals $A$ and $B$ with $x_1<x_2<x_3<x_4$.
$CFT^{\otimes mn}$ and $CFT^{\otimes m}$ are the product theory with the replica manifolds described before. $\sigma_g$s are the local twist operators at the endpoints of each interval. The operators $\sigma_{gA}$, $\sigma_{{gA}^{-1}}$, $\sigma_{gB}$ and $\sigma_{{gB}^{-1}}$ are defined by the connection between all replica sheets in the \emph{mn}-product theory, but the operators $\sigma_{gm}$ and $\sigma_{{gm}^{-1}}$ are defined just by the connection between \emph{m}-replica sheets. So it makes sense that when \emph{n} is equal to one, we have the following relation
\begin{equation}\label{sigN01}
  \sigma_{gA}=\sigma_{gB}=\sigma_{gm}.
\end{equation}

\section{Computation of the $S_R$ in the GMMG/GCFT at two disjoint intervals  }\label{sec:3}
In this section we consider the reflected entropy $S_R$ in the framework of GMMG/GCFT where the bulk geometry has an asymptotically flat geometry and the Galilean conformal field theory (GCFT) exist at the 2d boundary. The gravity describing the bulk geometry is the generalized minimal massive gravity (GMMG) \cite{GMMG01}.
The GMMG model has the following 3-form Lagrangian \cite{GMMG01}
\begin{equation}\label{GM01}
  L_{GMMG}=L_{TMG}-\frac{1}{m^2}\big(f.R+\frac{1}{2}e.f\times f\big)+\frac{\alpha}{2}e.h\times h
\end{equation}
 where $L_{TMG}$ and $R$ are the TMG Lagrangian and the dualised curvature, respectively. There are two auxiliary fields $f$ and $h$, that are one-form fields. $m$ is a mass parameter term, $\alpha$ is also a parameter and $e$  is a dreibein. The Chern-Simons (CS-) term in the TMG Lagrangian has the following form in the first order formalism
 \begin{equation}\label{CS01}
   CS-term: \frac{1}{2\mu}(\omega.d\omega+\frac{1}{3}\omega.\omega\times\omega)
 \end{equation}
where $\omega$ is dualised spin-connection and $\mu$ is the mass parameter of TMG. The algebra of the asymptotic conserved charges of asymptotically $AdS_3$ spacetimes in the context of GMMG is isomorphic to two copies of the Virasoro algebra with the following right and left central charges \cite{GMMG01, GMM03}
\begin{equation}\label{Vir.cen2}
  c_{GMMG}^+=\frac{3l}{2G}(-\sigma-\frac{\alpha H}{\mu}-\frac{F}{m^2}+\frac{1}{\mu l}),~~~c_{GMMG}^-=\frac{3l}{2G}(-\sigma-\frac{\alpha H}{\mu}-\frac{F}{m^2}-\frac{1}{\mu l}),
\end{equation}
where $H$ and $F$ are constant. By taking Inonu-Wigner contraction as \cite{GC01,GC02}, we can find the asymptotic symmetry group as a
 GCA with the following central charges
 \begin{eqnarray}\label{cLM-GMMG}
  c_L &=& c_{GMMG}^+-c_{GMMG}^-=\frac{3}{\mu G}, \nonumber \\
  c_M &=& \frac{(c_{GMMG}^++c_{GMMG}^-)}{l}=\frac{3}{G}(-\sigma-\frac{\alpha H}{\mu}-\frac{F}{m^2}).
\end{eqnarray}

To compute the $S_R$ in the GMMG/GCFT scenario, we consider the GCFT at the boundary in three different cases: the vacuum, a finite temperature and a finite sized system. In these cases, there are two disjoint intervals to describe the bipartite mixed configuration of the states at the boundary of the duality. The geometry of the bulk dual to the GCFT could be described in three different asymptotically flat spaces. The computations for each case are given in the following subsections.

\subsection{The disjoint intervals in vacuum}\label{sec:301}
In this subsection we compute the reflected entropy $S_R$ for the bipartite states with two disjoint intervals in the vacuum state of the $GCFT_2$ that is dual to the asymptotically flat geometry in the bulk described by 3d Poincare spacetime with the following metric
\begin{equation}\label{Poin01}
  ds^2=-du^2+dr^2+r^2d\phi^2,
\end{equation}
this is obtained from the Poincare patch of the AdS spacetime (when the AdS radius $\to \infty$) \cite{Po01}. So the metric in the Cartesian coordinates has the following form
\begin{equation}\label{Poin02}
  ds^2=-dt^2+dx^2+dz^2.
\end{equation}

We choose the intervals as follows
\begin{equation}\label{int01}
  A=[(x_1,t_1),(x_2,t_2)],~~~B=[(x_3,t_3),(x_4,t_4)].
\end{equation}

We have an intuitive computation on our previous work \cite{GMMG02} on the minimal EWCS in the GMMG/GCFT scenario in the configuration similar to the configuration mentioned in this subsection. The result of the EWCS in this bipartite mixed state is as follows
\begin{equation}\label{EW01}
  E_W = -\frac{1}{4G}(\sigma+\frac{\alpha H}{\mu}+\frac{F}{m^2})\big(\frac{x_{13}}{t_{13}}+\frac{x_{24}}{t_{24}}-\frac{x_{14}}{t_{14}}-\frac{x_{23}}{t_{23}}   \big) + \frac{1}{4\mu G}\log \frac{t_{13}t_{24}}{t_{14}t_{23}}.
\end{equation}

We use the replica manifold construction introduced in section \ref{sec:2} to express the proper form of the Renyi entropy and its related manifolds. To compute the Renyi entropy, we consider $Z_{n,m}$ in the numerator of $\exp\big((1-n)S_n(AA^*) \big)$ in (\ref{Rnm04}). Utilizing (\ref{Rnm05}), $Z_{n,m}$ can be described by 4-point correlation functions of the twist operators as follows
\begin{equation}\label{Z301}
  Z_{n,m}=\big\langle\sigma_{g_A}(x_1,t_1)\sigma_{{g_A}^{-1}}(x_2,t_2)
 \sigma_{g_B}(x_3,t_3)\sigma_{{g_B}^{-1}}(x_4,t_4)\big\rangle_{GCFT^{\otimes mn}},
\end{equation}

We are allowed to use the twist operators in writing the correlation functions of the primary fields in the Galilean conformal field theories \cite{GC04}.
The correlation functions in (\ref{Z301}) can be expanded in the t-channel using the $GCA_2$ conformal blocks $\mathcal{F}_{\alpha}$ as follows \cite{RE02,Hij}
\begin{eqnarray}\label{Z302}
   &\big\langle &\sigma_{g_A}(x_1,t_1)\sigma_{{g_A}^{-1}}(x_2,t_2)
 \sigma_{g_B}(x_3,t_3)\sigma_{{g_B}^{-1}}(x_4,t_4)\big\rangle_{GCFT^{\otimes mn}} \nonumber\\
   &=& \sum_{\alpha}\mathcal{F}_{\alpha}(c,h_p,\chi_p,h_i,\chi_i,x,t)
\end{eqnarray}
where $x$ and $t$ are the cross ratios for the $GCFT_2$ with the following definitions \cite{RE02,Hij}
\begin{eqnarray}\label{CR01}
  x &=& \frac{x_{12}x_{34}}{x_{13}x_{24}}, \nonumber\\
  t &=& \frac{t_{12}t_{34}}{t_{13}t_{24}}.
\end{eqnarray}

$(h_i,\chi_i)$ are the conformal weights of the twist fields $\sigma_{g_B}$ and $\sigma_{{g_A}^{-1}}$
and $(h_p,\chi_p)$ are the conformal weights of the primary twist operators $\sigma_{{g_B}{g_A}^{-1}}$.
$a_p$ can be defined by the coefficients in the OPE expansion as follows
\begin{equation}\label{OP01}
  a_p=c_{12}^pc_{34}^p.
\end{equation}
where $c^p_{ij}$ is an OPE coefficient. Using the result in \cite{RE01} and \cite{Seng03}, the conformal weights have the following relation
\begin{eqnarray}\label{hip01}
  h_i &=& h_{g_B}=h_{g_A^{-1}}=\frac{c_L}{24}n(m-\frac{1}{m}), \nonumber\\
  h_p &=& h_{g_Bg_A^{-1}}=\frac{c_L}{12}(n-\frac{1}{n}), \nonumber\\
  \chi_i &=& \chi_{g_B}=h_{g_A^{-1}}=\frac{c_M}{24}n(m-\frac{1}{m}), \nonumber\\
  \chi_p &=& \chi_{g_Bg_A^{-1}}=\frac{c_M}{12}(n-\frac{1}{n}).
\end{eqnarray}

The authors in \cite{OE,mon01,per01} have expanded the 4-point correlation functions of the twist operators in the $CFT^{\otimes mn}$ replica construction by the Virasoro conformal blocks. The relation of the Virasoro conformal blocks in the large central charge limit can be found in \cite{mon02,mon03}.
By using the result on the 4-point correlation functions of the twist operators in the $CFT^{\otimes mn}$ replica construction, we find the following result in the large central charge limit
\begin{eqnarray}\label{4pc01}
  &\log&\big\langle\sigma_{gA}(x_1)\sigma_{{gA}^{-1}}(x_2)
 \sigma_{gB}(x_3)\sigma_{{gB}^{-1}}(x_4)\big\rangle_{CFT^{\otimes mn}}  \nonumber \\
   &=& -4h_i \log [(x_4-x_1)(x_3-x_2)]-4h_p\log 2m \nonumber \\
   &+&4h_p \log 2-4h_p \log \frac{1+\sqrt{1-x}}{\sqrt{x}}
\end{eqnarray}
where by taking the limit $m\to 1$, we find the following result
\begin{eqnarray}\label{4pc02}
  &\log&\big\langle\sigma_{gA}(x_1)\sigma_{{gA}^{-1}}(x_2)
 \sigma_{gB}(x_3)\sigma_{{gB}^{-1}}(x_4)\big\rangle_{CFT^{\otimes mn}} \nonumber \\
   &=& -4h_i \log [(x_4-x_1)(x_3-x_2)]-4h_p \log \frac{1+\sqrt{1-x}}{\sqrt{x}}.
\end{eqnarray}

Since the $Z_{1,m}$ also plays a role in the Renyi entropy (\ref{Rnm05}), we obtain it by using the above relation (\ref{4pc02}) as follows
\begin{eqnarray}\label{4pc03}
  &\log&\big\langle\sigma_{gm}(x_1)\sigma_{{gm}^{-1}}(x_2)
 \sigma_{gm}(x_3)\sigma_{{gm}^{-1}}(x_4)\big\rangle_{CFT^{\otimes m}}\nonumber \\
   &=& -4\frac{h_i}{n} \log [(x_4-x_1)(x_3-x_2)],
\end{eqnarray}
where we have used the conformal weights (\ref{hip01}) in $n=1$. Then by substituting (\ref{4pc02}) and (\ref{4pc03}) into (\ref{Rnm05}), we find the following result
\begin{equation}\label{Rnm-c01}
  S_n(AA^*)=\frac{1}{1-n}\big\{-4h_p \log \frac{1+\sqrt{1-x}}{\sqrt{x}}   \big\},
\end{equation}
where in the $m,n\to 1$ limit, the reflected entropy for the two disjoint intervals at the $CFT_2$ can be found as follows
\begin{equation}\label{RAc01}
 S_R(A:B)\approx\frac{2c}{3}\log\bigg(\frac{1+\sqrt{1-x}}{\sqrt{x}}   \bigg)
\end{equation}
where we have used the $h_p$ relation (\ref{hip01}). This result is in consistent with the result in \cite{RE01}.

\subsection*{The general form of the reflected entropy in GCFT  }
The 4-point correlation functions of the twist operators in GCFT theories at the boundary of the duality have the following form \cite{GC02}
\begin{eqnarray}\label{4pg01}
  &\big\langle&\sigma_{gA}(x_1)\sigma_{{gA}^{-1}}(x_2)
 \sigma_{gB}(x_3)\sigma_{{gB}^{-1}}(x_4)\big\rangle_{GCFT^{\otimes mn}}  \nonumber \\
   &=& t^{-2h_i}t_{14}^{-2h_i}t_{23}^{-2h_i}\exp\big[-2\chi_i\frac{x_{23}}{t_{23}}
   -2\chi_i\frac{x_{14}}{t_{14}}-2\chi_i\frac{x}{t} \big]\mathcal{F}(t,\frac{x}{t}),
\end{eqnarray}
where in the t-channel ($t\to1$ and $x\to 0$), the 4-point correlations in the GCFT have the following relation
\begin{eqnarray}\label{4pg02}
  &\big\langle&\sigma_{gA}(x_1)\sigma_{{gA}^{-1}}(x_2)
 \sigma_{gB}(x_3)\sigma_{{gB}^{-1}}(x_4)\big\rangle_{GCFT^{\otimes mn}}  \nonumber \\
   &=& t_{14}^{-2h_i}t_{23}^{-2h_i}\exp\big[-2\chi_i\frac{x_{23}}{t_{23}}
   -2\chi_i\frac{x_{14}}{t_{14}} \big]\mathcal{F}(t,\frac{x}{t}),
\end{eqnarray}
where $\mathcal{F}(t,\frac{x}{t})$ is a sum over the Galilean conformal blocks that have the follwoing relation \cite{GC02}
\begin{equation}\label{CBg01}
  \mathcal{F}(t,\frac{x}{t})\equiv \sum_{p}\bigg(\frac{1+\sqrt{t}}{\sqrt{1-t}} \bigg)^{-2h_p}e^{-2\chi_p \frac{x}{2\sqrt{t}(t-1)}}
\end{equation}

The sum in (\ref{CBg01}) is over the $p$ states of the twist fields that are different from the $i$-states. The former are the intermediate or internal states in the OPE expansion and the latter are the external states in the OPE expansion. The conformal weights can be found in (\ref{hip01}). Using the result (\ref{4pg02}), the logarithm of the 4-point function is as follows
\begin{eqnarray}\label{4pg03}
  &\log&\big\langle\sigma_{gA}(x_1)\sigma_{{gA}^{-1}}(x_2)
 \sigma_{gB}(x_3)\sigma_{{gB}^{-1}}(x_4)\big\rangle_{GCFT^{\otimes mn}}  \nonumber \\
   &=& -2h_i \log [(t_4-t_1)(t_3-t_2)]-2\chi_i\big(\frac{x_{23}}{t_{23}}+\frac{x_{14}}{t_{14}}   \big)\nonumber\\
   &-& 2h_p \log\bigg(\frac{1+\sqrt{t}}{\sqrt{1-t}} \bigg)-2\chi_p\frac{x}{2\sqrt{t}(t-1)}.
\end{eqnarray}

Since the $Z_{1,m}$ also plays a role in the Renyi entropy (\ref{Rnm05}), we obtain it by using the relation (\ref{4pg03}) as follows
\begin{eqnarray}\label{4pg04}
  &\log&\big\langle\sigma_{gm}(x_1)\sigma_{{gm}^{-1}}(x_2)
 \sigma_{gm}(x_3)\sigma_{{gm}^{-1}}(x_4)\big\rangle_{GCFT^{\otimes m}}  \nonumber \\
   &=& -\frac{2h_i}{n} \log [(t_4-t_1)(t_3-t_2)]-\frac{2\chi_i}{n}\big(\frac{x_{23}}{t_{23}}+
   \frac{x_{14}}{t_{14}}\big).
\end{eqnarray}
where the $n$ is assumed to be equal to one and $h_i\to \frac{h_i}{n}$, $\chi_i\to \frac{\chi_i}{n}$ and $h_p=\chi_p=0$ in this case. Substituting the following relation into the (\ref{4pg03}),
\begin{eqnarray}\label{CBg02}
  f_L(t) \equiv \log\bigg(\frac{1+\sqrt{t}}{\sqrt{1-t}} \bigg), \nonumber\\
  f_M(t, \frac{x}{t})\equiv \frac{x}{2\sqrt{t}(t-1)},
\end{eqnarray}
we find the general form of the logarithm of the 4-point correlation function of the $GCFT^{\otimes mn}$ construction as follows
\begin{eqnarray}\label{4pg05}
  &\log&\big\langle\sigma_{gA}(x_1)\sigma_{{gA}^{-1}}(x_2)
 \sigma_{gB}(x_3)\sigma_{{gB}^{-1}}(x_4)\big\rangle_{GCFT^{\otimes mn}}  \nonumber \\
   &=& -2h_i \log [(t_4-t_1)(t_3-t_2)]-2\chi_i\big(\frac{x_{23}}{t_{23}}+\frac{x_{14}}{t_{14}}   \big)\nonumber\\
   &-& 2h_p f_L(t)-2\chi_p f_M(t, \frac{x}{t}).
\end{eqnarray}
where $f_L(t)$ and $f_M(t, \frac{x}{t})$ are considered as the conformal blocks related to $c_L$ and $c_M$, respectively. These conformal blocks should be calculated in each case of the flat holography configurations.
To find the general form of the Renyi for the bipartite state in the $GCFT^{\otimes mn}$ construction, we substitute (\ref{4pg04}) and (\ref{4pg05}) into (\ref{Rnm05}) and we find the follwoing result
\begin{equation}\label{RnG01}
  S_n(AA^*)=\frac{1}{1-n}\big(-2h_p f_L(t)-2\chi_p f_M(t, \frac{x}{t})     \big),
\end{equation}
where by taking the $m,n\to 1$ limit and using the conformal weights relations (\ref{hip01}), we find the general form of the reflected entropy for the flat holography as follows
\begin{equation}\label{Reg05}
  S_R(A:B)=\frac{c_L}{3}f_L(x,t)+\frac{c_M}{3}f_M(x,t),
\end{equation}
where the theory in the boundary should be the $GCFT_2$ with the bipartite mixed state configuration.
In this way, we have divided the reflected entropy into two parts that this result is consistent with the idea presented in \cite{RE06}. In that paper, the Renyi entropy derivative in theories without chiral symmetry is considered proportional to the sum of the left- and right- part of the Renyi entropy as follows
\begin{equation}\label{Reg03}
  \frac{\partial S_n}{\partial z_i}\propto c_L\frac{\partial f_L}{\partial z_i}+c_M\frac{\partial f_M}{\partial z_i}.
\end{equation}

The author in \cite{RE06} considers the Renyi entropy (\ref{Reg03}) in chiral symmetric theories with $c_L=c_M$ as follows
\begin{equation}\label{Reg04}
  \frac{\partial S_n}{\partial z_i}\propto 2c \frac{\partial f}{\partial z_i}.
\end{equation}

\subsection*{Return to the computation for the vacuum }
The authors in \cite{Seng02} have computed the conformal block for the two disjoint intervals in the vacuum of the $GCFT_2$ which is dual to the asymptotically flat geometry described by the Poincare metric (\ref{Poin02}) in TMG model. According to our intuitive computations in \cite{GMMG02}, we can use the same conformal block in the Poincare spacetime which is described by the GMMG.\footnote{In \cite{GMMG02},
we provide an argument
based on the holographic entanglement entropy. The GMMG model and the TMG model
are similar in some respects in that they include the Chern-Simons and the Einstein gravity
parts, but the former includes the higher derivative gravity terms. By comparing the results in \cite{GC05} and \cite{HEESK} for the entanglement entropy in flat-TMG/GCFT and flat-GMMG/GCFT scenarios respectively, we have found the effect of the differences between GMMG and
TMG is reflected only in the central charges in the result of the holographic entanglement
entropy. Then we made a conjecture that the same result could be true for the minimal EWCS, and in the transition from flat-TMG to flat-GMMG, the geometric calculations did not change and we just need to change the central charges. These geometric calculations include the extremal length of the surface is homologous to the entangling surface at the boundary and the extremized boost is needed to drag the normal vector of the particle frame through the trajectory of the spinning particle. Since it can be shown that these geometric calculations are proportional to the conformal blocks \cite{Hij02}, so we guess we can use the results of \cite{Seng02} for the Virasor conformal blocks needed to compute the reflected entropy in the context of the flat-GMMG/GCFT holography.   } So using the result in \cite{Seng02}, we write the $f_M(x,t)$ in (\ref{CBg02}) for the bipartite configuration with two disjoint intervals in the vacuum and in the t-channel ($t\to 1$) as follows
\begin{equation}\label{fM2d01}
  f_M=\frac{1}{2}\big(\frac{x_{13}}{t_{13}}+\frac{x_{24}}{t_{24}}-\frac{x_{14}}{t_{14}}-\frac{x_{23}}{t_{23}}     \big).
\end{equation}

The $f_L(x,t)$ in the reflected entropy is the conformal block which is obtained from the Chern-Simons (CS-) term in the TMG Lagrangian. Because of the similarity between the CS-term in the TMG and GMMG \cite{GMMG01}, we can use the result in \cite{Seng02} for the GMMG case. The $f_L(x,t)$ which is the conformal block related to the CS-term in the Lagrangian could have the following relation \cite{Seng02}
\begin{equation}\label{fL2d01}
  f_L=\frac{1}{2}\log\big(\frac{t_{13}t_{24}}{t_{14}t_{23}}   \big).
\end{equation}

Substituting (\ref{fM2d01}) and (\ref{fL2d01}) into (\ref{Reg05}) and using the central charges (\ref{cLM-GMMG}), the reflected entropy could have the following result
\begin{equation}\label{RE2d01}
  S_R(A:B)=\frac{1}{2\mu G}\log\big(\frac{t_{13}t_{24}}{t_{14}t_{23}}   \big)
  +\frac{1}{2G}(-\sigma-\frac{\alpha H}{\mu}-\frac{F}{m^2})\big(\frac{x_{13}}{t_{13}}+\frac{x_{24}}{t_{24}}-\frac{x_{14}}{t_{14}}-\frac{x_{23}}{t_{23}}     \big).
\end{equation}

This result is twice the result for minimal EWCS (\ref{EW01}) in our previous work \cite{GMMG02}. It should be noted, for the two disjoint intervals at the boundary, we find the results for the reflected entropy and the minimal EWCS  in proportion to the conformal block in the t-channel where $t\to 1$ and $x\to 0$. If the computations, both in the boundary and in the bulk side of the flat holography, are done in the s-channel, we get a trivial result. This is why in the papers \cite{RE02,Hij}, they work in t-channel for the computations of four-point functions in flat holography.

\subsection{The disjoint intervals at a finite temperature}\label{sec:302}
In this subsection, the bipartite states with two disjoint intervals $A$ and $B$ are considered in a thermal $GCFT_2$ on a cylinder with the temperature that is equal to its inverse circumference. The bulk geometry described by the GMMG can be given by flat space cosmology (FSC) in three dimensions as follows\cite{GC04,GC05,GC07}
\begin{equation}\label{FSC01}
  ds^2=Mdu^2-2du dr+r^2d\phi^2,
\end{equation}
where the $M$ parameter and the  $GCFT_2$ temperature have the following relation
\begin{equation}\label{M01}
  M=\big(\frac{2\pi}{\beta}  \big)^2.
\end{equation}

The authors in \cite{Seng02} have computed the conformal block for the two disjoint intervals in a thermal the $GCFT_2$ with finite temperature which is dual to the asymptotically flat space cosmology described by the FSC metric (\ref{FSC01}) in TMG model. According to results in \cite{GMMG02} and the argument provided in the footnote in previous subsection , we can use the same conformal block in the FSC spacetime which is described by the GMMG. So using the result in \cite{Seng02}, we write the conformal block $f_M(\tilde{x},\tilde{t})$ for the two disjoint intervals in a thermal $GCFT_2$ as follows
\begin{eqnarray}\label{fM2d02}
  f_M &=& \frac{\tilde{x}}{2\sqrt{\tilde{t}}(\tilde{t}-1)}, \nonumber\\
   &=&  \frac{1}{4}\bigg(\frac{\sqrt{M}u_{13}}{\tanh\frac{\sqrt{M}\phi_{13}}{2}}
  +\frac{\sqrt{M}u_{24}}{\tanh\frac{\sqrt{M}\phi_{24}}{2}}-\frac{\sqrt{M}u_{14}}{\tanh\frac{\sqrt{M}\phi_{14}}{2}}
  -\frac{\sqrt{M}u_{23}}{\tanh\frac{\sqrt{M}\phi_{23}}{2}}\bigg)
\end{eqnarray}
where $\tilde{x}$ and $\tilde{t}$ have the following relations \cite{Seng02}
\begin{eqnarray}\label{txt01}
  \tilde{t} &=& \frac{\sinh\frac{\sqrt{M}\phi_{12}}{2}\sinh\frac{\sqrt{M}\phi_{34}}{2}}
  {\sinh\frac{\sqrt{M}\phi_{13}}{2}\sinh\frac{\sqrt{M}\phi_{24}}{2}} , \nonumber\\
  \frac{\tilde{x}}{\tilde{t}} &=& \frac{1}{2}\bigg(\frac{\sqrt{M}u_{12}}{\tanh\frac{\sqrt{M}\phi_{12}}{2}}
  +\frac{\sqrt{M}u_{34}}{\tanh\frac{\sqrt{M}\phi_{34}}{2}}-\frac{\sqrt{M}u_{13}}{\tanh\frac{\sqrt{M}\phi_{13}}{2}}
  -\frac{\sqrt{M}u_{24}}{\tanh\frac{\sqrt{M}\phi_{24}}{2}}\bigg).
\end{eqnarray}
where the plane and the cylinder coordinates have the following relations
\begin{equation}\label{xt-pc}
  x=e^{i\phi},~~~t=iu e^{i\phi}.
\end{equation}

The $f_L(\tilde{x},\tilde{t})$ in the reflected entropy that is the conformal block related to the CS-term in the Lagrangian could have the following relation \cite{Seng02}
\begin{eqnarray}\label{fL2d02}
  f_L &=& \log\bigg(\frac{1+\sqrt{\tilde{t}}}{\sqrt{1-\tilde{t}}}   \bigg) \nonumber\\
  &=& \frac{1}{2}\log\bigg( \frac{\sinh\frac{\sqrt{M}\phi_{13}}{2}\sinh\frac{\sqrt{M}\phi_{24}}{2}}
  {\sinh\frac{\sqrt{M}\phi_{14}}{2}\sinh\frac{\sqrt{M}\phi_{23}}{2}}    \bigg)
\end{eqnarray}
as mentioned above, because of the similarity between the CS-term in the TMG and GMMG \cite{GMMG02}, it is permissible to use the result of the CS-term in the TMG Lagrangian in \cite{Seng02} in the GMMG case.

Substituting (\ref{fL2d02}) and (\ref{fM2d02}) into (\ref{Reg05}) and utilizing the central charges (\ref{cLM-GMMG}), the reflected entropy could have the following result
\begin{eqnarray}\label{RE2d02}
  S_R(A:B) &=& \frac{1}{2\mu G}\log\bigg( \frac{\sinh\frac{\sqrt{M}\phi_{13}}{2}\sinh\frac{\sqrt{M}\phi_{24}}{2}}
  {\sinh\frac{\sqrt{M}\phi_{14}}{2}\sinh\frac{\sqrt{M}\phi_{23}}{2}}    \bigg)\nonumber\\
  && +\frac{1}{4G}(-\sigma-\frac{\alpha H}{\mu}-\frac{F}{m^2})\bigg(\frac{\sqrt{M}u_{13}}{\tanh\frac{\sqrt{M}\phi_{13}}{2}}
  +\frac{\sqrt{M}u_{24}}{\tanh\frac{\sqrt{M}\phi_{24}}{2}} \nonumber\\
  &&-\frac{\sqrt{M}u_{14}}{\tanh\frac{\sqrt{M}\phi_{14}}{2}}
  -\frac{\sqrt{M}u_{23}}{\tanh\frac{\sqrt{M}\phi_{23}}{2}}\bigg).
\end{eqnarray}
This result is twice the result for minimal EWCS (\ref{EW01}) in our previous work \cite{GMMG02}.

\subsection{The disjoint intervals in a finite sized system}\label{sec:303}
In the following, the bipartite states with two disjoint intervals $A$ and $B$ are considered in a $GCFT_2$ compactified on a circle of the circumference $L_{\phi}$. The bulk geometry described by the GMMG can be given by the global
Minkowski orbifold in three dimensions as follows \cite{Po01}
\begin{equation}\label{glob01}
  ds^2=-(\frac{2\pi}{L_{\phi}})^2du^2-2du dr+r^2d\phi^2,
\end{equation}
where the subsystem size has the following relation
\begin{equation}\label{M02}
  M=-(\frac{2\pi}{L_{\phi}})^2.
\end{equation}

As we can see, the forms of the metrics in (\ref{FSC01}) and (\ref{glob01}) are similar except in the definition of the ADM mass parameter $M$ of the spacetime. Based on the claim made \cite{GMMG02}, we are allowed to use the TMG/GCFT results in the GMMG/GCFT flat holography to find the Virasoro conformal blocks needed in the reflected entropy relation. Substituting the relation $\sqrt{M}=\frac{2\pi i}{L_\phi}$ into the $\tilde{t}$ and $\tilde{x}$ relation (\ref{txt01}), we have the cross ratios as follows
\begin{eqnarray}\label{txt02}
  T &=& \frac{\sin\frac{\pi\phi_{12}}{L_{\phi}}\sin\frac{\pi\phi_{34}}{L_{\phi}}}
  {\sin\frac{\pi\phi_{13}}{L_{\phi}}\sin\frac{\pi\phi_{24}}{L_{\phi}}} , \nonumber\\
  \frac{X}{T} &=& \frac{\pi}{L_{\phi}}\bigg(\frac{u_{12}}{\tan\frac{\pi\phi_{12}}{2}}
  +\frac{u_{34}}{\tan\frac{\pi\phi_{34}}{L_{\phi}}}-\frac{u_{13}}{\tan\frac{\pi\phi_{13}}{L_{\phi}}}
  -\frac{u_{24}}{\tan\frac{\pi\phi_{24}}{L_{\phi}}}\bigg).
\end{eqnarray}

Substituting the cross ratios (\ref{txt02}) into the conformal block relations (\ref{fM2d02}) and (\ref{fL2d02}), we have the following relation for the reflected entropy for the bipartite state in a $GCFT_2$ on a compactified circle at the boundary as follows
\begin{eqnarray}\label{RE2d03}
  S_R(A:B) &=& \frac{1}{2\mu G}\log\bigg( \frac{\sin\frac{\pi\phi_{13}}{L_{\phi}}\sin\frac{\pi\phi_{24}}{L_{\phi}}}
  {\sin\frac{\pi\phi_{14}}{L_{\phi}}\sin\frac{\pi\phi_{23}}{L_{\phi}}}    \bigg)\nonumber\\
  && +\frac{\pi}{2G L_{\phi}}(-\sigma-\frac{\alpha H}{\mu}-\frac{F}{m^2})\bigg(\frac{u_{13}}{\tan\frac{\pi\phi_{13}}{L_{\phi}}}
  +\frac{u_{24}}{\tan\frac{\pi\phi_{24}}{L_{\phi}}} \nonumber\\
  &&-\frac{u_{14}}{\tan\frac{\pi\phi_{14}}{L_{\phi}}}
  -\frac{u_{23}}{\tan\frac{\pi\phi_{23}}{L_{\phi}}}\bigg).
\end{eqnarray}
where we have substituted the new relations for $f_L$ and $f_M$ into (\ref{Reg05}).

\section{Computation of the $S_R$ in the GMMG/GCFT at two adjacent intervals  }\label{sec:4}
In the previous section, we presented an approach to compute the reflected entropy for the bipartite state of the GCFT at the boundary of the flat holography. We considered the bulk geometry as an asymptotically flat spacetime described by the generalized minimal massive gravity (GMMG). In this approach, the reflected entropy $S_R(A:B)$ is obtained from Renyi entropy formula (\ref{Rnm04}), which is related to the 4-point correlation functions of the twist operators with a specific structure in (\ref{Rnm05}). Here, we are allowed to use the general form of the reflected entropy for the bipartite states as in (\ref{Reg05}). In this section, we extend the approach mentioned above for the bipartite states with two adjacent intervals of a GCFT at the boundary. Here, as in the previous section, we consider three different cases for the boundary and the bulk of the duality.

\subsection{ The adjacent intervals in vacuum} \label{sec:401}
As in the subsection \ref{sec:301}, we consider the two adjacent intervals $A=[(x_1,t_1),(x_2,t_2)]$ and
$B=[(x_3,t_3),(x_4,t_4)]$ in the vacuum state of the $GCFT_2$ which is dual to the asymptotically flat Minkowski spacetime with the metric (\ref{Poin01}). The GMMG can be considered as the gravity to describe the bulk geometry. By extending our conjecture \cite{GMMG02} on the minimal EWCS in the GMMG/GCFT flat holography to the adjacent intervals scenario, the minimal EWCS has the following relation
\begin{equation}\label{EWA01}
  E_W = \frac{1}{4G}(-\sigma-\frac{\alpha H}{\mu}-\frac{F}{m^2})\big(\frac{x_{12}}{t_{12}}+\frac{x_{23}}{t_{23}}-\frac{x_{13}}{t_{13}}   \big) + \frac{1}{4\mu G}\log \frac{t_{12}t_{23}}{\epsilon\big(t_{12}+t_{23}\big)}.
\end{equation}

To find the conformal blocks $f_M$ and $f_L$, we use the result in \cite{Seng02}. These conformal blocks have the following relations \cite{Seng02}
\begin{equation}\label{fM2a01}
  f_M(x,t)=\frac{1}{2}\bigg(\frac{x_{12}}{t_{12}}+\frac{x_{23}}{t_{23}}-\frac{x_{13}}{t_{13}}    \bigg)
\end{equation}
and
\begin{equation}\label{fL2a01}
  f_L(x,t)=\frac{1}{2}\log \frac{t_{12}t_{23}}{\epsilon\big(t_{12}+t_{23}\big)}.
\end{equation}

Substituting the conformal blocks (\ref{fM2a01}) and (\ref{fL2a01}) and the central charges (\ref{cLM-GMMG})  into the reflected entropy relation (\ref{Reg05}), we find the following result for the bipartite adjacent intervals in the GCFT vacuum that is dual to the GMMG by the Minkowski spacetime (\ref{Poin01})
\begin{equation}\label{RE2a01}
  S_R(A:B)=\frac{1}{2G}(-\sigma-\frac{\alpha H}{\mu}-\frac{F}{m^2})\big(\frac{x_{12}}{t_{12}}+\frac{x_{23}}{t_{23}}-\frac{x_{13}}{t_{13}}   \big) + \frac{1}{2\mu G}\log \frac{t_{12}t_{23}}{\epsilon\big(t_{12}+t_{23}\big)}.
\end{equation}
As expected, this result is twice the minimal EWCS (\ref{EWA01})

\subsection{ The adjacent intervals at a finite temperature} \label{sec:402}
The bipartite state configuration of the boundary with two adjacent intervals can be considered by the thermal $GCFT_2$ on a cylinder with the temperature that is equal to its inverse circumference. As in the subsection \ref{sec:302}, the GMMG is the gravity describing the bulk geometry can be given by the flat space cosmology (FSC) (\ref{FSC01}). The $M$ relation with the GCFT temperature in (\ref{M01}) can be used in this subsection. The adjacent intervals in the cylinder coordinates are as follows
\begin{equation}\label{AB02}
  A=\big[\big(u_1, \phi_1\big),\big(u_2, \phi_2   \big)\big],
  ~~~B=\big[\big(u_3, \phi_3\big),\big(u_4, \phi_4   \big)\big].
\end{equation}

The minimal EWCS in this configuration can be obtained by extending the conjecture on \cite{GMMG02} and utilizing the result in \cite{Seng02}. The minimal EWCS for the adjacent intervals in a thermal GCFT which is dual to the GMMG with the FSC geometry has the following relation
\begin{eqnarray}\label{EWA02}
  E_W &=& \frac{1}{4G}(-\sigma-\frac{\alpha H}{\mu}-\frac{F}{m^2})
  \bigg[\frac{\frac{\pi u_{12}}{\beta}}{\tanh\big(\frac{\pi\phi_{12}}{\beta}\big)}
  +\frac{\frac{\pi u_{23}}{\beta}}{\tanh\big(\frac{\pi\phi_{23}}{\beta}\big)}
  -\frac{\frac{\pi u_{13}}{\beta}}{\tanh\big(\frac{\pi\phi_{13}}{\beta}\big)}     \bigg] \nonumber\\
  &+& \frac{1}{4\mu G}\log\bigg[\frac{\beta}{\pi\epsilon}\frac{\sinh\big(\frac{\pi\phi_{12}}{\beta}\big)
  \sinh\big(\frac{\pi\phi_{23}}{\beta}\big)}
  {\sinh\big(\frac{\pi(\phi_{12}+\phi_{23})}{\beta}\big)}      \bigg].
\end{eqnarray}

The conformal blocks $f_M$ and $f_L$, can be found using the result in \cite{Seng02}. These conformal blocks for the two adjacent intervals in a thermal GCFT have the following relations \cite{Seng02}
\begin{equation}\label{fM2a02}
  f_M(x,t)=\frac{1}{2}\bigg[\frac{\frac{\pi u_{12}}{\beta}}{\tanh\big(\frac{\pi\phi_{12}}{\beta}\big)}
  +\frac{\frac{\pi u_{23}}{\beta}}{\tanh\big(\frac{\pi\phi_{23}}{\beta}\big)}
  -\frac{\frac{\pi u_{13}}{\beta}}{\tanh\big(\frac{\pi\phi_{13}}{\beta}\big)}     \bigg]
\end{equation}
and
\begin{equation}\label{fL2a02}
  f_L(x,t)=\frac{1}{2}\log\bigg[\frac{\beta}{\pi\epsilon}\frac{\sinh\big(\frac{\pi\phi_{12}}{\beta}\big)
  \sinh\big(\frac{\pi\phi_{23}}{\beta}\big)}
  {\sinh\big(\frac{\pi(\phi_{12}+\phi_{23})}{\beta}\big)}      \bigg].
\end{equation}

Substituting (\ref{fM2a02}) and (\ref{fL2a02}) and the central charges (\ref{cLM-GMMG})  into the reflected entropy relation (\ref{Reg05}), the reflected entropy in this configuration has the following result
\begin{eqnarray}\label{RE2a02}
  S_R(A:B) &=& \frac{1}{2G}(-\sigma-\frac{\alpha H}{\mu}-\frac{F}{m^2})
  \bigg[\frac{\frac{\pi u_{12}}{\beta}}{\tanh\big(\frac{\pi\phi_{12}}{\beta}\big)}
  +\frac{\frac{\pi u_{23}}{\beta}}{\tanh\big(\frac{\pi\phi_{23}}{\beta}\big)}
  -\frac{\frac{\pi u_{13}}{\beta}}{\tanh\big(\frac{\pi\phi_{13}}{\beta}\big)}     \bigg]\nonumber\\
  &+& \frac{1}{2\mu G}\log\bigg[\frac{\beta}{\pi\epsilon}\frac{\sinh\big(\frac{\pi\phi_{12}}{\beta}\big)
  \sinh\big(\frac{\pi\phi_{23}}{\beta}\big)}
  {\sinh\big(\frac{\pi(\phi_{12}+\phi_{23})}{\beta}\big)}      \bigg].
\end{eqnarray}
in which the bipartite adjacent intervals are in the GCFT vacuum which is dual to the Minkowski flat spacetime (\ref{Poin01}) described by the GMMG. In this section as well, as expected, the result (\ref{RE2a02}) is twice the minimal EWCS (\ref{EWA02})

\subsection{ The adjacent intervals in a finite sized system} \label{sec:403}
In this subsection, two adjacent intervals $A$ and $B$ are considered in a $GCFT_2$ compactified on a circle of the circumference $L_{\phi}$. The bulk geometry described by the GMMG can be given by the 3d global
Minkowski orbifold (\ref{glob01}). The relation between $M$ and $L_{\phi}$ in (\ref{M02}) determines the similarity between the metrics in (\ref{FSC01}) and (\ref{glob01}). Substituting the relation $\beta\to i L_{\phi}$ into the minimal EWCS (\ref{EWA02}), we have the following relation for the minimal EWCS for the two adjacent intervals in a $GCFT_2$ on a compactified circle at the boundary of the duality as follows
\begin{eqnarray}\label{EWA03}
  E_W &=& \frac{1}{4G}(-\sigma-\frac{\alpha H}{\mu}-\frac{F}{m^2})
  \bigg[\frac{\frac{\pi u_{12}}{L_{\phi}}}{\tan\big(\frac{\pi\phi_{12}}{L_{\phi}}\big)}
  +\frac{\frac{\pi u_{23}}{L_{\phi}}}{\tan\big(\frac{\pi\phi_{23}}{L_{\phi}}\big)}
  -\frac{\frac{\pi u_{13}}{L_{\phi}}}{\tan\big(\frac{\pi\phi_{13}}{L_{\phi}}\big)}     \bigg] \nonumber\\
  &+& \frac{1}{4\mu G}\log\bigg[\frac{L_{\phi}}{\pi\epsilon}\frac{\sin\big(\frac{\pi\phi_{12}}{L_{\phi}}\big)
  \sin\big(\frac{\pi\phi_{23}}{L_{\phi}}\big)}
  {\sin\big(\frac{\pi(\phi_{12}+\phi_{23})}{L_{\phi}}\big)}      \bigg].
\end{eqnarray}

 The conformal blocks $f_M$ and $f_L$  for the two adjacent intervals in a compactified GCFT have the following relations \cite{Seng02}
\begin{equation}\label{fM2a03}
  f_M(x,t)=\frac{1}{2}\bigg[\frac{\frac{\pi u_{12}}{L_{\phi}}}{\tan\big(\frac{\pi\phi_{12}}{L_{\phi}}\big)}
  +\frac{\frac{\pi u_{23}}{L_{\phi}}}{\tan\big(\frac{\pi\phi_{23}}{L_{\phi}}\big)}
  -\frac{\frac{\pi u_{13}}{L_{\phi}}}{\tan\big(\frac{\pi\phi_{13}}{L_{\phi}}\big)}     \bigg]
\end{equation}
and
\begin{equation}\label{fL2a03}
  f_L(x,t)=\frac{1}{2}\log\bigg[\frac{L_{\phi}}{\pi\epsilon}\frac{\sin\big(\frac{\pi\phi_{12}}{L_{\phi}}\big)
  \sin\big(\frac{\pi\phi_{23}}{L_{\phi}}\big)}
  {\sin\big(\frac{\pi(\phi_{12}+\phi_{23})}{L_{\phi}}\big)}      \bigg].
\end{equation}

Substituting (\ref{fM2a03}) and (\ref{fL2a03}) and the central charges (\ref{cLM-GMMG})  into the reflected entropy relation (\ref{Reg05}), the reflected entropy in this configuration has the following result

\begin{eqnarray}\label{RE2a03}
  S_R(A:B) &=& \frac{1}{2G}(-\sigma-\frac{\alpha H}{\mu}-\frac{F}{m^2})
  \bigg[\frac{\frac{\pi u_{12}}{L_{\phi}}}{\tan\big(\frac{\pi\phi_{12}}{L_{\phi}}\big)}
  +\frac{\frac{\pi u_{23}}{L_{\phi}}}{\tan\big(\frac{\pi\phi_{23}}{L_{\phi}}\big)}
  -\frac{\frac{\pi u_{13}}{L_{\phi}}}{\tan\big(\frac{\pi\phi_{13}}{L_{\phi}}\big)}   \bigg] \nonumber\\
  &+& \frac{1}{2\mu G}\log\bigg[\frac{L_{\phi}}{\pi\epsilon}\frac{\sin\big(\frac{\pi\phi_{12}}{L_{\phi}}\big)
  \sin\big(\frac{\pi\phi_{23}}{L_{\phi}}\big)}
  {\sin\big(\frac{\pi(\phi_{12}+\phi_{23})}{L_{\phi}}\big)}      \bigg].
\end{eqnarray}

The result (\ref{RE2a03}) for the reflected entropy is twice the result (\ref{EWA03}) for the minimal EWCS, as expected.

\section{Conclusion  }\label{sec:5}
In this paper, we considered a class of the reflected entropy as a new correlation measure for the bipartite mixed states. The reflected entropy has been introduced in \cite{RE01} to
be a boundary dual to the minimal entanglement wedge cross section (EWCS) and is proportional to the algorithmically constructed region in the bulk. The result that the reflected entropy is twice as great as the minimal EWCS is proved in the AdS/CFT holography in \cite{RE01}. . In this paper, we utilized the replica technique developed in \cite{RE01} in the flat holography case where we considered the two dimensional Galilean
conformal field theory ($GCFT_2$) at the boundary of the holography. The bulk geometry of the flat holography was explained by the generalized massive gravity (GMMG) that is a theory without the bulk-boundary clash as mentioned in the introduction.

To find the reflected entropy for the bipartite state in the GCFT at the boundary, we explained an appropriate construction of the Renyi entropy utilizing the replica technique which is used in the related papers \cite{RE01,RE02}. The $m,n\to1$ limit of the Renyi entropy results in the bipartite reflected entropy in (\ref{Rnm05}) and the 4-point correlation functions of the twist operators are related to the reflected entropy. So finding the correlation functions is important. We considered the correlation functions in details in both CFT and GCFT construction. The conformal blocks are the dominant terms of the correlation function expansions in the t-channel ($T\to 1$ and $X\to 0$). We found the general form of the reflected entropy (\ref{Reg05}) has two different parts that each of them corresponds to a part of the conformal weights of the twist operators. This is in consistence with the lack of chiral symmetry in the GMMG and the GCFT. The first part is related to the Einstein gravity and the higher derivative terms in the GMMG Lagrangian and the second part is related to the Chern-Simons (CS-) modification in the Lagrangian.  To find the conformal blocks, we referred to the paper \cite{Seng02}, which is related to the TMG/GCFT flat holography. What convinced us to use the results of \cite{Seng02} were the intuitive computations we did in the previous work \cite{GMMG02}. In \cite{GMMG02}, we came to the conclusion that given the differences between the TMG and the GMMG, we are allowed to use the calculations of the former conformal blocks in the latter as well. Using the conformal blocks relations into the reflected entropy (\ref{Reg05}), we obtained the reflected entropy for two disjoint intervals in three different cases in (\ref{RE2d01}), (\ref{RE2d02}) and (\ref{RE2d03}) and for two adjacent intervals in the related cases in (\ref{RE2a01}), (\ref{RE2a02}) and (\ref{RE2a03}). In all of these relations, we show the relation (\ref{RE01}) between the reflected entropy and minimal EWCS could be a true one.

The results we have obtained are important in several aspects. First, we have been able to generalize the related computations from the AdS/CFT holography to a class of the non-AdS holography. Second, the relation between the reflected entropy and the minimal EWCS is also confirmed in this class of the holography, so given the difficulty of the purification computations, we can still consider the reflected entropy as a proper measure for the classical and the quantum correlations. Third, the Renyi entropy construction presented in this type of the computations is more appropriate than the same construction for the entanglement entropy, and the relation between the correlation functions and the conformal blocks is well expressed. Of course, we know very well that for the bipartite mixed states configurations of the boundary of the holography, the entropy entanglement is not a proper measure for the entanglement. Finally, the good accordance we obtained between the reflected entropy and minimal EWCS in the GMMG/GCFT holograpy, was a good sign of correctness of the intuitive computations we did in the previous paper \cite{GMMG02}.

\section*{Acknowledgements}
We would like to thank Mohammad Hassan Vahidinia for discussions and comments.


\begin{thebibliography}{99}

\bibitem{hol01}
G. t Hooft, \emph{Dimensional reduction in quantum gravity} ,\text{gr-qc/9310026} .

\bibitem{hol02}
L. Susskind, \emph{The world as a hologram}, \emph{J. Math. Phys.} {\bf 36} (1995) 6377.

\bibitem{hol03}
J.M. Maldacena, \emph{The large N limit of superconformal field theories and supergravity}, \emph{Int. J.
Theor. Phys.} 38 (1999) 1113.


\bibitem{hol04}
S.S. Gubser, I.R. Klebanov and A.M. Polyakov, \emph{Gauge theory correlators from noncritical
string theory}, \emph{Phys. Lett. B} 428 (1998) 105.

\bibitem{hol05}
E. Witten, \emph{Anti-de Sitter space and holography}, \emph{Adv. Theor. Math. Phys.} 2 (1998) 253.


\bibitem{EE03}
M. Levin and X.-G. Wen, \emph{Detecting Topological Order in a Ground State Wave Function},
\emph{Phys. Rev. Lett.} 96 (2006) 110405.


\bibitem{EE04}
H. Casini and M. Huerta, \emph{A Finite entanglement entropy and the c-theorem}, \emph{Phys. Lett. B}
600 (2004) 142.

\bibitem{EE05}
H. Casini and M. Huerta, \emph{On the RG running of the entanglement entropy of a circle}, \emph{Phys.
Rev. D} 85 (2012) 125016.


\bibitem{EE08}
S. Ryu and T. Takayanagi, \emph{Holographic derivation of entanglement entropy from AdS/CFT},
Phys. Rev. Lett. 96 (2006) 181602.

\bibitem{EE09}
S. Ryu and T. Takayanagi, \emph{Aspects of Holographic Entanglement Entropy}, \emph{JHEP} 08 (2006)
045.

\bibitem{EE10}
V.E. Hubeny, M. Rangamani and T. Takayanagi, \emph{A Covariant holographic entanglement
entropy proposal}, \emph{JHEP} 07 (2007) 062.

\bibitem{EW01}
M. Headrick, V. E. Hubeny,  A. Lawrence, and M. Rangamani, \emph{Causality and holographic entanglement entropy}, \emph{JHEP} 2014, 162 (2014).

\bibitem{EW02}
D. L. Jafferis and S. J. Suh, \emph{The Gravity Duals of Modular Hamiltonians}, (2014), \emph{JHEP} 09 (2016) 068.

\bibitem{EW03}
D. L. Jafferis, A. Lewkowycz, J. Maldacena, \emph{Relative entropy equals bulk relative entropy}, and S. J. Suh, \emph{JHEP} {\bf 6}, 4 (2016).

\bibitem{CW01}
A. Almheiri, X. Dong, and D. Harlow, \emph{Bulk Locality and Quantum Error Correction in AdS/CFT}, \emph{JHEP} 2015, 163 (2015).

\bibitem{CW02}
F. Pastawski, B. Yoshida, D. Harlow, and J. Preskill, \emph{Holographic quantum error-correcting codes: toy models for the bulk/boundary correspondence}, \emph{JHEP} 2015, 149 (2015).

\bibitem{CW03}
V. E. Hubeny and M. Rangamani, \emph{Causal Holographic Information}, \emph{JHEP} 2012, 114 (2012).

\bibitem{EP01}
T. Takayanagi and K. Umemoto, \emph{Entanglement of purification through holographic
duality}, \emph{Nature Phys.} {\bf 14} no. 6, (2018) 573–577.

\bibitem{EP02}
P. Nguyen, T. Devakul, M. G. Halbasch, M. P. Zaletel, and B. Swingle, \emph{Entanglement of
purification: from spin chains to holography}, \emph{JHEP} {\bf 01} (2018) 098.

\bibitem{EP03}
A. Bhattacharyya, T. Takayanagi, and K. Umemoto, \emph{Entanglement of Purification in Free Scalar Field Theories}, \emph{JHEP} 2018, 132 (2018).

\bibitem{EP04}
H. Hirai, K. Tamaoka, and T. Yokoya, \emph{Towards Entanglement of Purification for Conformal Field Theories}, \emph{Progress of Theoretical and Experimental Physics} 2018, 063B03 (2018).
\bibitem{EP05}
B. M. Terhal, M. Horodecki, D. W. Leung, and D. P. DiVincenzo, \emph{The entanglement of
purification}, \emph{J. Math. Phys.} 43 (2002) 4286.

\bibitem{EP06}
K. Umemoto and Y. Zhou, \emph{Entanglement of Purification for Multipartite States and its
Holographic Dual}, \emph{JHEP} 10 (2018) 152.

\bibitem{OE}
K. Tamaoka, \emph{Entanglement Wedge Cross Section from the Dual Density Matrix}, \emph{Phys. Rev. Lett.} {\bf 122} 14, (2019) 141601.

\bibitem{RE01}
S. Dutta and T. Faulkner, \emph{A canonical purification for the entanglement wedge
cross-section}, \emph{JHEP} 03 (2021) 178.

\bibitem{RE02}
H.-S. Jeong, K.-Y. Kim, and M. Nishida, \emph{Reflected Entropy and Entanglement Wedge
Cross Section with the First Order Correction}, \emph{JHEP} 12 (2019) 170.

\bibitem{RE03}
N. Bao and N. Cheng, \emph{Multipartite Reflected Entropy}, \emph{JHEP} 10 (2019) 102.

\bibitem{RE04}
J. Chu, R. Qi, and Y. Zhou, \emph{Generalizations of Reflected Entropy and the Holographic
Dual}, \emph{JHEP} 03 (2020) 151.


\bibitem{Ku01}
J. Kudler-Flam and S. Ryu, \emph{Entanglement negativity and minimal entanglement wedge cross sections in holographic theories}, \emph{Phys. Rev. D} {\bf 99} (2019) 106014.


\bibitem{GC01}
A. Bagchi and R. Gopakumar, \emph{Galilean Conformal Algebras and AdS/CFT}, \emph{JHEP} 07 (2009) 037.


\bibitem{GC02}
A. Bagchi, R. Gopakumar, I. Mandal and A. Miwa, \emph{GCA in 2d}, \emph{JHEP} 08 (2010) 004.

\bibitem{GC03}
A. Bagchi and I. Mandal, \emph{On Representations and Correlation Functions of Galilean
Conformal Algebras}, \emph{Phys. Lett. B} {\bf 675} (2009) 393-397.

\bibitem{GC04}
R. Basu and M. Riegler, \emph{Wilson Lines and Holographic Entanglement Entropy in Galilean
Conformal Field Theories}, \emph{Phys. Rev. D} {\bf 93} (2016) 045003.

\bibitem{holG01}
G. Barnich and C. Troessaert, \emph{Aspects of the BMS/CFT correspondence}, \emph{JHEP} 05 (2010)
062.


\bibitem{GMMG01}
M. R. Setare, \emph{On the generalized minimal massive gravity}, \emph{Nucl. Phys. B} 898 (2015)
259-275.

\bibitem{MMG01}
E. Bergshoeff, O. Hohm, W. Merbis, A. J. Routh and P. K. Townsend, \emph{Class. Quant. Grav.} {\bf 31}, 145008 (2014).

\bibitem{TMG01}
S. Deser, R. Jackiw and S. Templeton, \emph{Three-Dimensional Massive Gauge Theories},
\emph{Phys. Rev. Lett.} {\bf 48} (1982) 975-978.

\bibitem{TMG02}
S. Deser, R. Jackiw and S. Templeton, \emph{Topologically Massive Gauge Theories}, \emph{Annals
Phys.} {\bf 140} (1982) 372-411.

\bibitem{NMG}
E. A. Bergshoeff, O. Hohm and P. K. Townsend, \emph{Phys. Rev. Lett.} 102, 201301 (2009).


\bibitem{mon01}
T. Hartman, \emph{Entanglement Entropy at Large Central Charge}, 1303.6955.


\bibitem{Seng01}
V. Malvimat, H. Himanshu, B. Paul and G. Sengupta, \emph{Entanglement Negativity in Galilean Conformal Field Theories}, \emph{Phys.Rev.D} 100 (2019) 2, 026001.



\bibitem{Seng02}
D. Basu, A. Chandra, V. Raj and G. Sengupta, \emph{Entanglement Wedge in Flat Holography and Entanglement Negativity}, arXiv:2106.14896.


\bibitem{GMMG02}

M. R. Setare and M. Koohgard, \emph{A conjecture on the minimal entanglement wedge cross section in the GMMG/GCFT flat holography }- to be appear.


\bibitem{RE05}
H-S. Jeong, K-Y. Kim and M. Nishida, \emph{Reflected entropy and entanglement wedge cross
section with the first order correction}, \emph{JHEP } 12 (2019) 170.


\bibitem{Hors}
R. Horodecki, P. Horodecki, M. Horodecki and K. Horodecki, \emph{Quantum entanglement}, \emph{Rev. Mod. Phys.} {\bf 81} (2009) 865.









\bibitem{CFT01}
P. Calabrese and J.L. Cardy, \emph{Entanglement entropy and quantum field theory}, \emph{J. Stat. Mech.}
0406 (2004) P06002.


\bibitem{CFT02}
P. Calabrese and J. Cardy, Entanglement entropy and conformal field theory, J. Phys. A 42
(2009) 504005.

\bibitem{GMM03}
M. R. Setare and H. Adami, \emph{Enhanced asymptotic BMS3 algebra of the flat spacetime solutions of generalized minimal massive gravity}, \emph{Nucl. Phys. B} {\bf 926} (2018) 70-82.

\bibitem{Po01}
G. T. Horowitz and R. Steif, \emph{Singular string solutions with nonsingular initial data},
\emph{Phys.Lett.B} 258 (1991) 91-96.


\bibitem{Hij}
E. Hijano, Semi-classical bms3 blocks and flat holography, \emph{JHEP} 2018 (Oct, 2018).

\bibitem{Seng03}
D. Basu, A. Chandra, H. Parihar and G. Sengupta, \emph{Entanglement Negativity in Flat
Holography}, arXiv:2102.05685.


\bibitem{per01}
E. Perlmutter, \emph{Virasoro conformal blocks in closed form}, \emph{JHEP} 08 (2015) 088.










\bibitem{mon02}
A. A. Belavin, A. M. Polyakov and A. B. Zamolodchikov, \emph{Infinite Conformal
Symmetry in Two-Dimensional Quantum Field Theory}, \emph{Nucl. Phys. B} 241, 333
(1984).

\bibitem{mon03}
Al. B. Zamolodchikov, \emph{Conformal symmetry in two-dimensional space: Recursion representation of conformal block}, \emph{Theor Math Phys} 73, 1088–1093 (1987).


\bibitem{RE06}
E. Perlmutter, \emph{Comments on Rényi entropy in} AdS3/CFT2 . \emph{JHEP.} 2014, 52 (2014).



\bibitem{GC05}
A. Bagchi, R. Basu, D. Grumiller and M. Riegler, \emph{Entanglement entropy in Galilean
conformal field theories and flat holography}, \emph{Phys. Rev. Lett.} {\bf 114} (2015) 111602.

\bibitem{HEESK}
M. R. Setare and M. Koohgard, \emph{Holographic entanglement entropy in flat limit of the generalized minimal massive gravity}, \emph{Eur.Phys.J.C.} 81 (2021) 8, 765.


\bibitem{Hij02}
E. Hijano and C. Rabideau, \emph{Holographic entanglement and Poincare blocks in
three-dimensional flat space}, \emph{JHEP} 05 (2018) 068.


\bibitem{GC07}
A. Bagchi and R. Basu, \emph{3D Flat Holography: Entropy and Logarithmic Corrections}, \emph{JHEP}
03 (2014) 020.








\end{thebibliography}
\end{document}